\documentclass[useAMS,usenatbib]{mn2e}

\usepackage{graphicx}  
\usepackage{dcolumn}   
\usepackage{bm}        
\usepackage{amsfonts,amsmath,amssymb,mathrsfs}
\usepackage{color}
\usepackage{mathrsfs}

\usepackage{hyperref}
\hypersetup{
  colorlinks=true,        
  linkcolor=blue,         
  citecolor=cyan,         
}
\usepackage{natbib,times}

\newcommand{\cf}{cf.~}
\newcommand{\ie}{i.e.,~}
\newcommand{\eg}{e.g.,~}

\title[Characterization of a black hole shadow]{A coordinate-independent characterization of a black hole
  shadow}
\author[ Abdujabbarov, Rezzolla $\&$ Ahmedov]
{A. A. Abdujabbarov,$^{1,2,3}$ 
L. Rezzolla$^{4,5}$ and 
B. J. Ahmedov$^{1,2,3}$%
\\ 
$^{1}$Institute of Nuclear Physics, Ulughbek, Tashkent 100214,
Uzbekistan\\
$^{2}$Ulugh Beg Astronomical Institute, Astronomicheskaya 33, Tashkent
100052, Uzbekistan\\
$^{3}$National University of Uzbekistan, Tashkent
100174, Uzbekistan\\
$^{4}$Institut f{\"u}r Theoretische Physik, Max-von-Laue-Str. 1, D-60438
Frankfurt, Germany\\
$^{5}$Frankfurt Institute for Advanced Studies, Ruth-Moufang-Str. 1, D-60438
Frankfurt, Germany}
\begin{document}

\date{\today}

\pagerange{\pageref{firstpage}--\pageref{lastpage}} 

\maketitle

\label{firstpage}

\begin{abstract}
A large international effort is under way to assess the presence of a
shadow in the radio emission from the compact source at the centre of our
Galaxy, Sagittarius A$^*$ (Sgr A$^*$). If detected, this shadow would
provide the first direct evidence of the existence of black holes and
that Sgr A$^*$ is a supermassive black hole. In addition, the shape of
the shadow could be used to learn about extreme gravity near the event
horizon and to determine which theory of gravity better describes the
observations. The mathematical description of the shadow has so far used
a number of simplifying assumptions that are unlikely to be met by the
real observational data. We here provide a general formalism to describe
the shadow as an arbitrary polar curve expressed in terms of a Legendre
expansion. Our formalism does not presume any knowledge of the properties
of the shadow, \eg the location of its centre, and offers a number of
routes to characterize the distortions of the curve with respect to
reference circles. These distortions can be implemented in a coordinate
independent manner by different teams analysing the same data. We show
that the new formalism provides an accurate and robust description of
noisy observational data, with smaller error variances when compared to
previous approaches for the measurement of the distortion.
\end{abstract}


\section{Introduction}
\label{sec:intro}

There is a widespread belief that the most convincing evidence for the
existence of black holes will come from the direct observations of
properties related to the horizon. These could be through the detection
of gravitational waves from the collapse to a rotating
star~\citep{Baiotti06}, from the ringdown in a binary black hole
merger~\citep{Berti:2009kk}, or through the direct observation of its
`shadow'. In a pioneering study, \citet{Bardeen73} calculated the shape
of a dark area of a Kerr black hole, that is, its `shadow' over a
bright background appearing, for instance, in the image of a bright star
behind the black hole. The shadow is a gravitationally lensed image of
the event horizon and depends on the closed orbits of photons around the
black hole\footnote{Strictly speaking, also an horizon-less object such
  as a gravastar~\citep{Mazur2004} would lead to a shadow. However,
  rather exotic assumptions on the heat capacity of the gravastar's
  surface are needed to justify a lack of emission from such a surface
  \citep{Broderick2007}; gravitational-wave emission would unambiguously
  signal the presence of an event
  horizon~\citep{chirenti_2007_htg}.}. Its outer boundary, which we will
hereafter simply refer to as the shadow, corresponds to the apparent
image of the photon capture sphere as seen by a distant observer. General
relativity predicts, in fact, that photons circling the black hole
slightly inside the boundary of the photon sphere will fall down into the
event horizon, while photons circling just outside will escape to
infinity. The shadow appears therefore as a rather sharp boundary between
bright and dark regions and arises from a deficit of those photons that
are captured by the event horizon. Because of this, the diameter of the
shadow does not depend on the photons' energy, but uniquely on the
angular momentum of the black hole. In general relativity and in an
idealized setting in which everything is known about the emission
properties of the plasma near the black hole, the shadow's diameter
ranges from $4.5\, r_{_S}$ for an extreme Kerr black hole, to
$\sqrt{27}\,r_{_S}$ for a Schwarzschild black hole, where $r_{_S} :=
2GM/c^2$ is the Schwarzschild radius. In practice, however, the size and
shape of the shadow will be influenced by the astrophysical properties of
the matter near the horizon and, of course, by the theory of gravity
governing the black hole.

Besides providing evidence on the existence of black holes, the
observation of the black hole shadow and of the deformations resulting in
the case of nonzero spin, is also expected to help determine many of the
black hole properties~\citep[see e.g. ][]{Chandrasekhar98, Falcke2013b,
    Takahashi04,Falcke2000,Doeleman2009}. More specifically, imaging the
shadow of a black hole via radio observations will allow one to test the
predictions of general relativity for the radius of the shadow and study
astrophysical phenomena in the vicinity of black holes [see
  \citet{Johannsen2010} and, more recently, \citet{Zakharov14} for a
  review]. In addition, it will allow one to set constraints on the
validity of alternative theories of gravity which also predict black
holes and corresponding shadows  \citep[see e.g.][]{Eiroa14, Tsukamoto14,
    Falcke2013b}.

The possible observation of a black hole shadow has recently received a
renewed attention as the spatial resolution attainable by very long
baseline interferometry (VLBI) radio observations is soon going to be
below the typical angular size of the event horizon of candidate
supermassive black holes (SMBHs), such as the one at the centre of the
Galaxy or the one in the M87 galaxy \citep{Falcke2000}. These
observations are the focus of international scientific collaborations,
such as the Event Horizon Telescope
(EHT)\footnote{http://eventhorizontelescope.org/} or the Black Hole
Camera (BHC)\footnote{http://blackholecam.org/}, which aim at VLBI
observations at 1.3 mm and 0.87 mm of Sagittarius A$^*$ (Sgr A$^*$) and
M87. We recall that Sgr A$^*$ is a compact radio source at the centre of
the Galaxy and the SMBH candidate in our galaxy. In fact, the orbital
motion of stars near Sgr A$^*$ indicates that its mass is $\simeq 4.3
\times 10^{6} M_\odot$~\citep{Ghez:2008,Genzel10}.

Given a distance of 8 kpc from us, the angular size of the Schwarzschild
radius of the SMBH candidate in Sgr A$^*$ is $\sim 10\,{\rm \mu as}$, so
that the corresponding angular diameter of the shadow is of the order of
$\sim 50\, {\rm \mu as}$. Similarly, with an estimated mass of $\simeq
6.4 \times 10^{9} M_\odot$ \citep{Gebhardt11} and a distance of 16 Mpc,
the M87 galaxy represents an equally interesting SMBH candidate, with an
angular size that is of the same order \ie $\sim 40\, {\rm \mu
  as}$. Although the resolution achievable at present is not sufficient
to observe an image of the shadow of either black hole, it is
sufficiently close that it is realistic to expect that near-future
observations will reach the required resolution. Indeed, future EHT and
BHC observations of Sgr A$^*$ are expected to go below the horizon scale
and to start to provide precise information on the black hole
orientation, as well as on the astrophysical properties of the accretion
flow taking place on to the black hole ~\citep{Chan15,Psaltis15}.

An extensive literature has been developed to calculate the shadow of the
black hole in known space-times, either within general relativity
\citep{Young76, Perlick04, Abdujabbarov2013}, or within alternative
theories of gravity \citep{Bambi2013, Tsukamoto14, Amarilla10,
  Amarilla12, Amarilla13, Atamurotov13a, Atamurotov13b, Schee09}. In most
cases, the expression of the shadow as a closed polar curve is not known
analytically, but for the Pleba\'nski-Demia\'nski class of space-times,
the shadow has been cast in an analytic
form~\citep{Grenzebach14,Grenzebach15}.

Because the shadow is in general a complex polar curve on the celestial
sky, an obvious problem that emerges is that of the characterization of
its deformation. For example, in the case of a Kerr black hole, the
difference in the photon capture radius between corotating and
counter-rotating photons creates a ``dent'' on one side of the shadow,
whose magnitude depends on the rotation rate of the black hole. A way to
measure this deformation was first suggested by \citet{Hioki09} and then
further developed by other authors~\citep{Bambi2009,Bambi10, Bambi2013}. In
essence, in these approaches the shape of the shadow is approximated as a
circle with radius $R_s$ and such that it crosses through three points
located at the poles and at the equator of the shadow's boundary. The
measure of the dent is then made in terms of the so called
`deflection', that is, the difference between the endpoints of the
circle and of the shadow, with a dimensionless distortion parameter being
given by the ratio of the size of the dent to the radius $R_s$ [\cf
  Eq. (\ref{delthioki})].

While this approach is reasonable and works well for a black hole such as
the Kerr black hole, it is not obvious it will work equally well for
black holes in more complex theories of gravity or even in arbitrary
metric theories of gravity as those considered by \citet{Rezzolla2014}.
Leaving aside the fact in all these works the shadow is assumed to be
determined with infinite precision (an assumption which is obviously
incompatible with a measured quantity), many but not all approaches in
characterizing the black hole shadow and its deformations suffer from at
least three potential difficulties. They often assume a primary shape,
\ie that the shadow can be approximated with a circle; exceptions are the
works of \citet{Kamruddin2013} and \citet{Psaltis2014}. In the first one,
a model has been proposed to describe the ``crescent'' morphology
resulting from the combined effects of relativistic Doppler beaming and
gravitational light bending \citep{Kamruddin2013}; in the second one, an
edge-detection scheme and a pattern-matching algorithm are introduced to
measure the properties of the black hole shadow even if the latter has an
arbitrary shape \citep{Psaltis2014}. \textit{(ii)} They assume that the
observer knows the exact position of the centre of black
hole\footnote{The reconstruction procedures of the observational data does
  resolve this problem, which is however still present when considering
  and comparing purely theoretical representations of the
  shadow}. \textit{(iii)} They are restricted to a very specific measure
of the distortion and are unable to model arbitrary distortion;
exceptions are the works of \citet{Johannsen2010} and
\citet{Johannsen2013}, where instead polar-averaged distortions have been
proposed.

To counter these potential difficulties, we present here a new general
formalism that is constructed to avoid the limitations mentioned
above. In particular, we assume that the shadow has an arbitrary shape
and expand it in terms of Legendre polynomials in a coordinate system
with origin in the effective centre of the shadow. This approach gives us
the advantage of not requiring the knowledge of the centre of the black
hole and of allowing us to introduce a number of parameters that measure
the distortions of the shadow. These distortions are both accurate and
robust, and can be implemented in a coordinate independent manner by
different teams analysing the same noisy data.

The paper is organized as follows. In Sect.~\ref{general:form} we develop
the new coordinate-independent formalism, where an arbitrary black hole
shadow is expanded in terms of Legendre polynomials. Using this
formalism, we introduce in Sect.~\ref{dist_params} various distortion
parameters of the shadow. In Sect.~\ref{metriccc} we apply the formalism
to a number of black hole space-times by computing the coefficients of the
expansion and by showing that they exhibit an exponentially rapid
convergence. We also compare the properties of the different distortion
parameters and assess which definition appears to be more accurate and
robust in general. Section~\ref{comparison} offers a comparison between
the new distortion parameters introduced here with the more traditional
ones simulating the noisy data that are expected from the
observations. Finally, Sect.~\ref{conclusion} summarizes our main results
the prospects for the use of the new formalism.

We use a system of units in which $c = G = 1$, a space-like signature
$(-,+,+,+)$ and a spherical coordinate system $(t,r,\theta,\phi)$. Greek
indices are taken to run from 0 to 3.

\section{General characterization of the shadow}
\label{general:form}

In what follows we develop a rather general formalism to describe the
black hole shadow that radio astronomical observations are expected to
construct. For all practical purposes, however, we will consider the
problem not to consist of the determination of the innermost unstable
circular orbits for photons near a black hole. Rather, we will consider
the problem of characterizing in a mathematically sound and
coordinate-independent way a closed curve in a flat space, as the one in
which the image will be available to us as distant observers.

\begin{figure}
\begin{center}
 \includegraphics[width=0.9\linewidth]{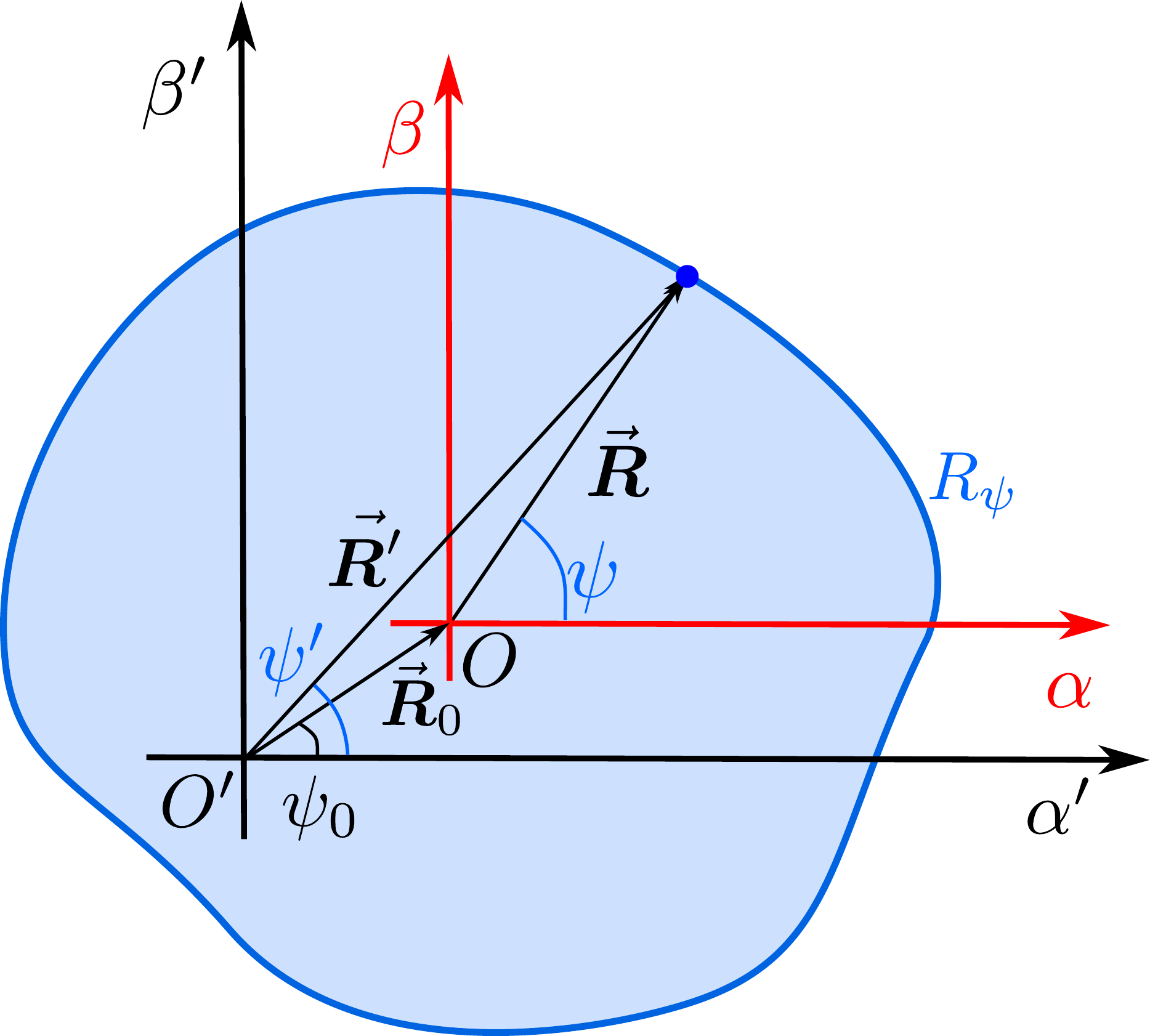}
\end{center}
\caption{Schematic representation of the black hole shadow as a generic
  polar curve $R_{\psi}$ in a coordinate system $(\alpha,\beta)$ with
  origin $O$ in the `centre' of the shadow. The latter is translated by
  a vector $\vec{\boldsymbol{R}}_{0}$ with respect to the arbitrary
  coordinate system $(\alpha',\beta')$ with origin $O'$ in which the
  observations are made.}
\label{cartoon_OO'}
\end{figure}

Assume therefore that the astronomical observations provide the shadow as
an one-dimensional closed curve defined by the equation
\begin{equation}
R'=R'(\psi')\,,
\end{equation}
where $R'$ and $\psi'$ can be thought of as the radial and angular
coordinates in a polar coordinate system with origin in $O'$. In
practice, astronomical observations will not be able to provide such a
sharp closed line and a more detailed analysis would need to take the
observational uncertainties (which could well be a function of $\psi'$)
into account. We will discuss some of these uncertainties in
Sect. \ref{comparison}, but for the time being we will consider the
shadow as an idealized one-dimensional closed curve. A schematic example
of the polar curve is shown in Fig.~\ref{cartoon_OO'}, where $\alpha'$
and $\beta'$ are the so-called ``celestial coordinates'' of the observer,
and represent an orthogonal coordinate system with one of the unit
vectors being along the line of sight.

Of course, there is no reason to believe that such a coordinate system is
particularly useful, or that in using it a nonrotating black hole will
have a shadow given by a perfect circle. Hence, in order to find a better
coordinate system, and, in particular, one in which a Schwarzschild black
hole has a circular shadow, we define the \textit{effective centre} of
the curve extending the definition of the position of the centre of mass
for a solid body to obtain
\begin{equation}
\label{eq:center_cont}
\vec{\boldsymbol{R}}_0 := 
\frac{
\int_0^{2\pi}
\vec{\boldsymbol{e}}_{_{R'}} (\psi')R' 
\left[g_{R'R'}(dR'/d\psi')^2 + g_{\psi'\psi'}\right]^{1/2}
\ d\psi'}
{\int_0^{2\pi} 
\left[g_{R'R'}(dR'/d\psi')^2 + g_{\psi'\psi'}\right]^{1/2}
\ d\psi'} \,,
\end{equation}
where $\vec{\boldsymbol{e}}_{_{R'}}$ is the radial-coordinate unit vector
and where $g_{R'R'}$, $g_{\psi'\psi'}$ are the metric functions of the
polar coordinate system $(R',\psi')$. Two important remarks: first, radio
observations may well yield, especially in the nearest future, only a
portion of the shadow, namely, the one with the largest brightness. Yet,
it is useful to consider here the shadow as a closed polar curve since
this is the way it is normally discussed in purely theoretical
investigations. Second, at least observationally, the location of the
centre of the black hole shadow is a free parameter in the
image-reconstruction procedure and so already part of the analysis of the
observational data. At the same time, the definition of a centre is
useful also in the absence of actual observational data since it can help
in the comparison of shadows that are built analytically and hence
without observational data.

From the knowledge of the vector $\vec{\boldsymbol{R}}_0$, the coordinate
position of the effective centre can be expressed explicitly in terms of
the radial and angular coordinates as
\begin{align}
\label{rzero}
&R_0 := \left(\int_0^{2 \pi}R'd\psi'\right)^{-1}
\Bigg[ \left(\int_0^{2\pi} R'^2 \cos\psi' d
  \psi'\right)^2\nonumber\\
&\hskip 2.5cm +
\left(\int_0^{2\pi} R'^2 \sin \psi' d \psi'\right)^2\Bigg]^{1/2}  \,,\\
\label{psizero}
&\psi_0 := \tan^{-1} \left(\frac{\int_0^{2\pi} R'^2 \sin \psi' d \psi'}
{\int_0^{2\pi} R'^2 \cos\psi' d \psi'}\right)\,.
\end{align}

We note that if the centre of the primary coordinate system $O'$
coincides with the black hole origin, then the parameter $R_0$ exactly
corresponds to the shift of the centre of the shadow with respect to the
black hole position defined by \citet{Tsukamoto14}.

Having determined the effective centre of the shadow, it is convenient to
define a new polar coordinate system centred in it with coordinates $(R,
\psi)$. Clearly, the new coordinate system with origin $O$ is just
translated by $\vec{\boldsymbol{R}}_{0}$ with respect to $O'$ and, hence,
the relation between the two coordinate systems is given by
\begin{align}
R :=& \Big[\left(R'\cos\psi'-R_0\cos\psi_0\right)^2\nonumber\\
&\hskip 1.0cm
+\left(R'\sin\psi'-R_0\sin\psi_0\right)^2\Big]^{1/2} \,,\\
\psi :=& \tan^{-1}\frac{R'\sin\psi'-R_0\sin\psi_0}
{R'\cos\psi'-R_0\cos\psi_0}\,.
\end{align}
Note that we have kept the new axes $\alpha$ and $\beta$ parallel to the
original ones $\alpha'$ and $\beta'$. This is not strictly necessary but
given the arbitrarity of the orientation of both sets of axes, it
provides a useful simplification.

A well-defined centre of coordinates allows us now to obtain a robust
definition of the reference \textit{areal circle} as the circle having
the same area as the one enclosed by the shadow. In particular, given the
closed parametric curve $R=R(\psi)$, its area will in general be given by
\begin{align}
{\mathcal A} := &
\int_{\psi_1}^{\psi_2} d\psi 
\int_{0}^{R} \sqrt{g_{\bar{R}\bar{R}}\, g_{\psi \psi}}  d\bar{R} \nonumber \\
= & \frac{1}{2}\int_{\psi_1}^{\psi_2} R^2d\psi =
\frac{1}{2} \int^{\lambda_2}_{\lambda_1} R^2(\lambda)
\left(\frac{d\psi}{d\lambda} \right)\, d\lambda \,,
\label{surface2}
\end{align}
where in the second equality we have set $g_{RR}=1$, $g_{\psi \psi}=R(\psi)$,
while in the third equality we consider the representation of the curve
in terms of a more generic parameter $\lambda$, \ie
$R=R(\psi(\lambda))$. If the shadow is a closed curve, the integration
limits $\lambda_{1,2}$ can be found from the condition $\psi(\lambda)=0$
and $\psi(\lambda)=2\pi$, while they will be restricted by the actual
observational data when only a portion of the shadow is available.

We can then define the \textit{areal radius} $R_{_{\mathcal A}}$ of the
reference circle simply as
\begin{equation}
R_{_{\mathcal A}} := \left(\frac{A}{\pi}\right)^{1/2}\,.
\end{equation}
Similarly, and if simpler to compute, it is possible to define the
\textit{circumferential radius} $R_c$ of the reference circle as
\begin{equation}
R_{_\mathcal{C}}  := \frac{\mathcal{C}}{2 \pi}\,,
\end{equation}
where the circumference is calculated as
\begin{align}
\mathcal{C} := &
\int \left(
dR^2 + g_{\psi\psi}\, d\psi^2\right)^{1/2}
\nonumber \\
= & 
\int_{\lambda_1}^{\lambda_2}\left[\left(\frac{dR}{d\lambda}\right)^2 +
R^2\left(\frac{d\psi}{d\lambda}\right)^2\right]^{1/2} d\lambda\,.
\end{align}

\begin{figure}
\begin{center}
 \includegraphics[width=0.9\linewidth]{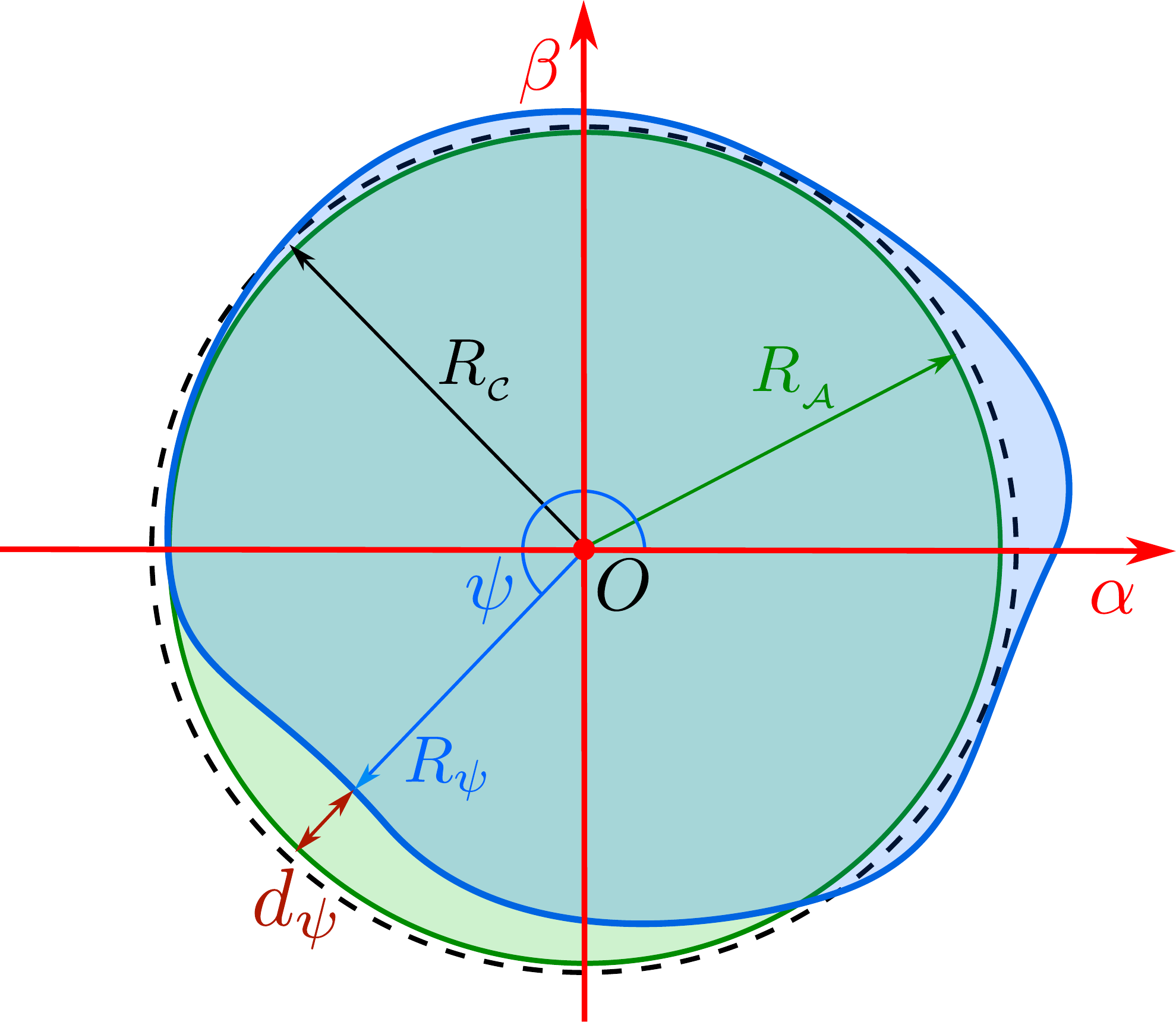}
\end{center}
\caption{Schematic representation of the local distortion $d_{\psi}$
  between the polar curve $R_{\psi}$ representing the black hole shadow
  and representative circles with circumference and area radii
  $R_{_\mathcal{C}}$ and $R_{_{A}}$, respectively.}
 \label{fig2}
\end{figure}

An areal radius is particularly useful as it enables one to measure two
useful quantities, namely, the local deviation of the shadow
$R_\psi:=R(\psi)$ from the areal circle, \ie 
\begin{equation}
d_\psi := |R_{_{\mathcal A}}-R_\psi| \,,
\end{equation}
and its polar average
\begin{equation}
d_{\langle \psi \rangle} :=
\frac{1}{2\pi}\int_0^{2\pi} d_\psi \, d\psi =
\frac{1}{2\pi}\int_0^{2\pi} |R_{_{\mathcal A}}-R_\psi|\, d\psi \,.
\end{equation}
Note that although similar, the areal and the circumferential radii are
in general different and coincide just for a spherically symmetric black
hole, in which case $R_{_{\mathcal A}} = R_c = R_{\psi}$, and of course
$d_{\psi}=0=d_s$. All of these geometrical quantities are shown
schematically in Fig.~\ref{fig2}.

\section{Distortion parameters}
\label{dist_params}

With a well defined and unambiguous set of coordinates $(R,\psi)$, we can
next move to the characterization of the geometrical properties of the
shadow. To this scope we simply employ an expansion in terms of Legendre
polynomial, \ie we define
\begin{equation}
\label{leg_exp}
R_{\psi} := \sum_{\ell=0}^{\infty} c_\ell P_{\ell}(\cos\psi)\,,
\end{equation}
where $P_{\ell}(\cos\psi)$ is the Legendre polynomial of order $\ell$ and
the coefficients $c_{\ell}$ of the expansion (\ref{leg_exp}) can be
found as
\begin{align}
c_{\ell} :=& \frac{2\ell+1}{2}\int_{0}^{\pi} R(\psi)
P_{\ell}(\cos\psi) \sin\psi\,  d\psi
\nonumber \\
=& \frac{2\ell+1}{2}
\int_{\lambda_1}^{\lambda_2} R(\psi)
P_{\ell}(\cos\psi) \sin\psi  \left(\frac{d\psi}{d\lambda}\right)
d\lambda \,.
\label{coef2}
\end{align}
The integration limits $\lambda_{1,2}$ can be found from the condition
$\psi(\lambda)=0$ and $\psi(\lambda)=\pi$, respectively. Using this
decomposition, it is straightforward to measure the differences between
the value of the parametrized shadow at two different angles. For
example, the relative difference between the shadow at $\psi=0$ and at a
generic angle $\psi=\pi/m$ can be computed simply as
\begin{align}
\label{eq:delta_n_m}
\delta_{m} :=& \frac{R_{\psi}(\psi=0) -
  R_{\psi}(\psi=\pi/m)}{R_{\psi}(\psi=0)} \nonumber \\
=& 1-\frac{\sum_{\ell=0}^{\infty} c_{\ell}
P_{\ell}(\cos\psi)|_{\psi=\pi/m}}{\sum_{\ell=0}^{\infty} c_{\ell}
P_{\ell}(\cos\psi)|_{\psi=0}}\,.
\end{align}
When $m=1$, expression~(\ref{eq:delta_n_m}) simplifies to
\begin{equation}
\delta_{1} := 1-\frac{\sum_{\ell=0}^{\infty} (-1)^{\ell}c_{\ell}}
{\sum_{\ell=0}^{\infty} c_{\ell} }\,,
\end{equation}
while, when $m=2$, the difference can still be computed analytically and
is given by
\begin{equation}
\delta_{2} := 1 - \frac{\mathscr{B}}{\mathscr{A}}
\,,
\end{equation}
where we have introduced the following and more compact notation that
will be used extensively in the remainder
\begin{align}
\mathscr{A} :=& R_{\psi}(\psi = 0) = \sum_{\ell=0}^{\infty} c_{\ell}\,,\\
\mathscr{B} :=& R_{\psi}(\psi = \pi/2) = \sum_{\ell=0}^{\infty}
(-1)^{\ell} \frac{(2\ell)! }{2^{2\ell} (\ell !)^2} c_{2\ell} \,,\\
\mathscr{C} :=& R_{\psi}(\psi = 3\pi/2) =
\sum_{\ell=0}^{\infty} (-1)^{\ell} c_{\ell}\,.
\end{align}
Analytic expressions for (\ref{eq:delta_n_m}) when $m > 2$ are much
harder to derive, but can be easily computed numerically.

We note that the parametrization~(\ref{eq:delta_n_m}) is quite general
and allows us to recover in a single definition some of the expressions
characterizing the distortion of the shadow and that have been introduced
by other authors. For example, the parameter $\delta_{n}^{1}$ can be
associated to the distortion parameter $\delta$ first introduced
by~\citet{Hioki09}~\citep[\cf Fig.~3 of][]{Hioki09}. Similarly, the parameter
$\delta_{4}$ is directly related to the distortion parameter $\epsilon$
introduced by~\citet{Tsukamoto14} \citep[\cf Fig.~3 of ][]{Tsukamoto14}.

In what follows we will exploit the general expression for the polar
curve representing the black hole shadow to suggest three different
definitions that measure in a coordinate-independent manner the amount of
distortion of the shadow relative to some simple background curve, \eg a
circle. These expressions are all mathematically equivalent and the use
of one over the other will depend on the specific properties of the
observed shadow.

\subsection{Distortion parameter -- {\rm I}}
\label{dp_I}

\begin{figure}
\begin{center}
\includegraphics[width=8cm]{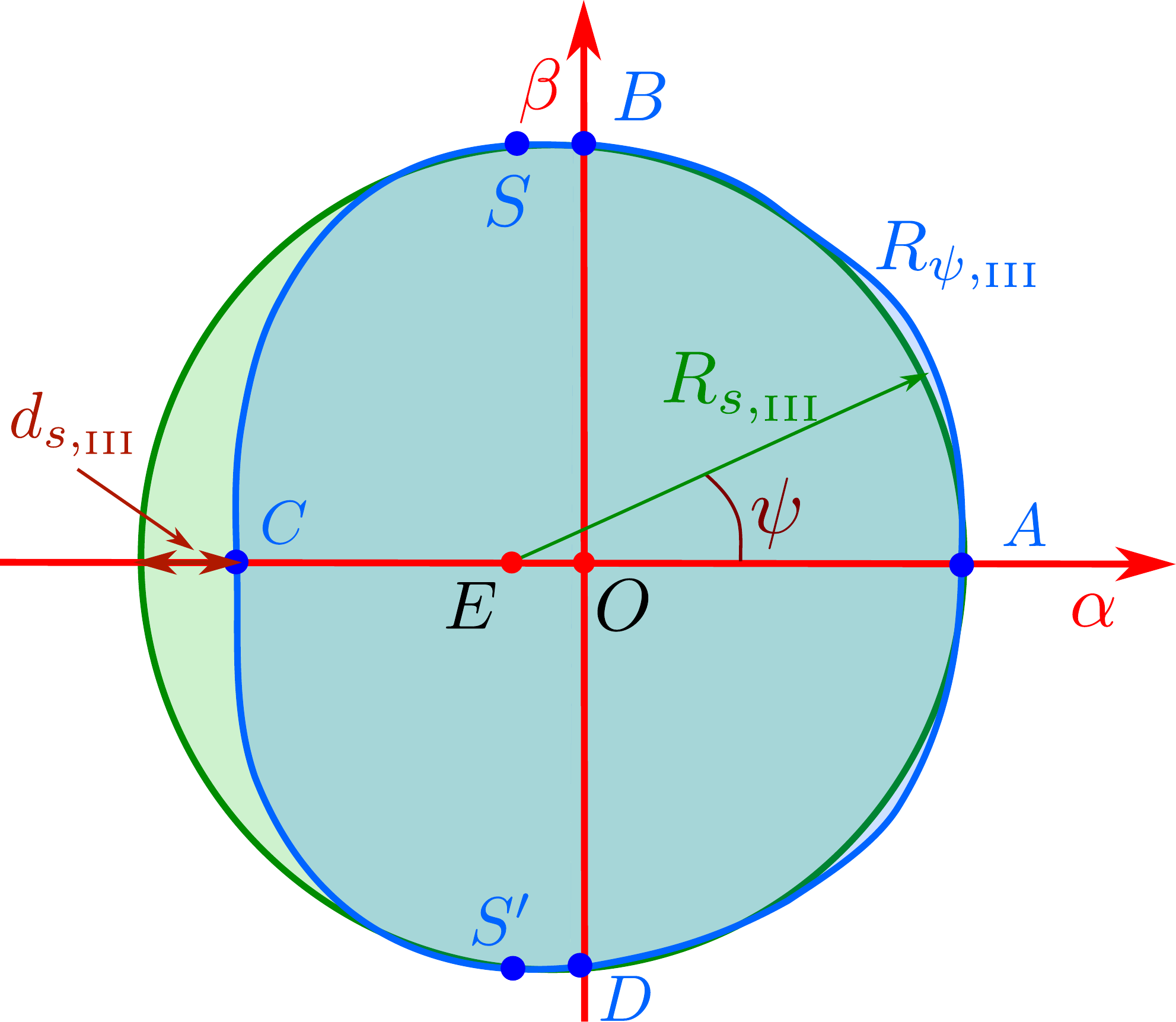}
\caption{Schematic representation of the distortion parameter -- III. The
  quantity $d_{s,_{\rm III}}$ measures the difference between the
  Legendre expanded polar curve $R_{\psi,_{\rm III}}$ and the reference
  circle of radius $R_{s,_{\rm III}}$ and passing through the points $A$,
  $B$, $D$, and having centre in point $E$. Also shown are the
  ``zero-slope'' points $S$ and $S'$.}
\label{fig3n}
\end{center}
\end{figure}

We start by considering three points on the polar curve $A$, $B$, and $D$,
which occupy precise angular positions at $\psi=0,\,\pi/2$, and $3\pi/2$,
respectively (see diagram in Fig.~\ref{fig3n}). The corresponding
distances $OA$, $OB$ and $OD$ from the centre of coordinates $O$ can then
be expressed as
\begin{align}
\label{OA}
\!\!\!
R_{_A} :=&  R_{\psi}(\psi =0) = \sum_{\ell=0}^{\infty} c_\ell
P_\ell(\cos\psi)|_{\psi=0}= \mathscr{A}\,, \\
\label{OB}
R_{_B} :=&  R_{\psi}(\psi = \pi/2) = \sum_{\ell=0}^{\infty} c_\ell
P_\ell(\cos\psi)|_{\psi=\pi/2} = \mathscr{B} \,,\ \ \\
\label{OD}
R_{_D} :=&  R_{\psi}(\psi = 3\pi/2) =  R_{_B} \,,
\end{align}
where in the last equality we have exploited the fact that the expansion
in Legendre polynomials is symmetric with respect to the $\alpha$ axis.

Next, we define a new parametric curve for which $R_{_A}=R_{_B}=R_{_D}$
and thus that satisfies the following condition
\begin{equation}
\label{sum_cls}
\mathscr{B} = \mathscr{A}\,,
\end{equation}
or, equivalently, for $\ell > 0$
\begin{equation}
  c_{2\ell-1} =   c_{2\ell} \left[ (-1)^\ell \frac{(2\ell)! }{2^{2\ell} (\ell !)^2}
  -1\right]\,.
\end{equation}

The corresponding polar expression, formulated in terms of the Legendre
polynomials expansion (\ref{leg_exp}), is therefore given by
\begin{align}
\label{polcurve_I}
&R_{\psi,_I}(\psi) =c_0+\sum_{\ell=1}^{\infty}c_{2\ell-1} \times \nonumber\\
&\left\{  P_{2\ell-1}
(\cos\psi) +
\left[(-1)^\ell \frac{(2\ell)!
}{2^{2\ell} (\ell !)^2} -1\right]^{-1} P_{2\ell} (\cos\psi)\right\},.
\nonumber \\
\end{align}

To measure the distortion we need a reference curve, which we can choose
to be the circle passing through the three points $A$, $B$, and $D$ and
thus with radius
\begin{equation}
\label{eq:ra.eq.rb}
R_{s,_{\rm I}} := R_{_A} = \mathscr{B} = R_{_B} = \mathscr{A}\,.
\end{equation}

We can now compute the deviation of the parametric curve
(\ref{polcurve_I}) from the corresponding background circle of radius
$R_{s,_{\rm I}}$ at any angular position. However, as customary in this
type of considerations, we can consider the shadow to be produced by a
rotating black hole with spin axis along the $\beta$ axis, so that the
largest deviations will be on the axis of negative $\alpha$ (this is
shown schematically in the left-hand panel of Fig.~\ref{fig:appendix}, when
considering the case of a Kerr black hole). More specifically, we can
define the difference between the curves at $\psi=\pi$ as
\begin{align}
d_{s,_{\rm I}} := R_{s,_{\rm I}} -R_{\psi,_I}(\psi=\pi)
=& \mathscr{B} - \sum_{\ell=0}^{\infty} c_{2\ell}+\sum_{\ell=1}^{\infty}
c_{2\ell-1}\nonumber\\
=& 2  \sum_{\ell=1}^{\infty} c_{2\ell-1}\,.
\end{align}
It follows that our first definition for the dimensionless distortion
parameter -- $\delta_{s,_{\rm I}}$ can then be given by
\begin{equation}
\label{deltadef1}
\delta_{s,_{\rm I}} := \frac{d_{s,_{\rm I}}}{R_{s,_{\rm I}}}=
\frac{2\sum_{\ell=1}^\infty c_{2\ell-1}}{\mathscr{B}}
=: \sum_{\ell=1}^\infty \delta_{\ell,_I} \,,
\end{equation}
which reduces to the compact expression
\begin{equation}
\delta_{s,_{\rm I}} \simeq \frac{2c_1}{c_0} = \delta_{1,_{\rm I}}\,.
\end{equation}
when only the first two coefficients in the expansion are taken into
account, \ie for $c_0\neq0\neq c_1$ and $c_\ell=0$, with $\ell \geq 2$
(we recall that $\delta_{0,_I}=0$). We also note that the assumption
\eqref{eq:ra.eq.rb} does not restrict the analysis to spherically
symmetric black hole space-times and, as we will show in
Fig. \ref{dist_allinone}, the distortion parameter \eqref{deltadef1} can
be applied also to axisymmetric space-times.

\subsection{Distortion parameter -- {\rm II}}
\label{dp_II}

A second possible definition of the distortion parameter is slightly more
general and assumes that the radial distance of points $A$ and $B$ from
the centre of coordinates is not necessarily the same, \ie $R_{_A} \neq
R_{_B}$. In this case, one can think of introducing a new point $E$ on
the $\alpha$ axis (this is shown schematically in the middle panel of
Fig.~\ref{fig:appendix}, when considering the case of a Kerr black hole),
such that the distances $AE=EB$ and which could therefore serve as the
centre of the reference circle. Since the values of the coordinates
$R_{_A}$ and $R_{_B}$ are defined by expressions (\ref{OA}) and
(\ref{OB}), we can use the condition $AE=EB$ to find that one can easily
find position of the point $E$ on the $\alpha$ axis is given by
\begin{equation}
R_{_E} = \left|\frac{R_{_B}^2-R_{_A}^2}{2 R_{_A}} \right| \,, \label{retwo}
\end{equation}
with the corresponding angular position $\psi_{_E}$ being either $0$ or
$\pi$, \ie
\begin{equation}
\psi_{_E} = \cos^{-1} \left({\frac{R_{_A}-R_{_B}}{|R_{_A}-R_{_B}|}}\right)\,.
\end{equation}
The radius of the circle passing through the three points $A$, $B$,
and $D$ is
\begin{equation}
R_{s,_{\rm II}}  =  \frac{R_{_B}^2+R_{_A}^2}{2 R_{_A}} \,,
\end{equation}
so that the deviation of the shadow at $\psi=\pi$ from the circle of
radius $R_{s,_{\rm II}}$ can be found using the relation
\begin{equation}
d_{s,_{\rm II}} = 2 R_{s,_{\rm II}} -(R_{_A}+R_{_C})\,.
\label{definitiondist}
\end{equation}

Finally, we can introduce the distortion parameter $d_{s,_{\rm II}}$ defined
as
\begin{equation}
d_{s,_{\rm II}} := \frac{\mathscr{B}^2}{\mathscr{A}}- \mathscr{C}\,,
\end{equation}
so that the second dimensionless distortion parameter is expressed as
\begin{equation}
\label{deltadef2}
\delta_{s,_{\rm II}} := \frac{d_{s,_{\rm II}}}{R_{s,_{\rm II}}} =
2\left(\frac{\mathscr{B}^2-\mathscr{A}\,\mathscr{C}}
{\mathscr{B}^2+\mathscr{A}^2}\right)\,.
\end{equation}

The expression for the dimensionless distortion (\ref{deltadef2}) is in
this case more complex that the one presented in Eq.~(\ref{deltadef1});
however, in the simpler case in which only the lowest order coefficients
are retained, \ie if $c_0\neq0\neq c_1$ and $c_\ell=0$ for $\ell \geq 2$,
we have
\begin{equation}
\label{deltas_II_trunc}
\delta_{s,_{\rm II}} \simeq \frac{2c_1^2}{2c_0^2+2c_0c_1+c_1^2}
= \delta_{1,_{\rm II}}\,.
\end{equation}

\subsection{Distortion parameter -- {\rm III}}
\label{dp_III}

A third and possibly \emph{optimal} definition of the distortion
parameter is one that is meant to consider the case in which the shadow
is still reflection symmetric relative to the $\alpha$ axis, but does not
cross the $\beta$ axis with a zero slope. Rather, the curve admits a
point, say $S$, at angular position $0 < \psi_{_S} < \pi$, where it has
zero slope relative to the $(\alpha, \beta)$ coordinate system (see
diagram in Fig.~\ref{fig3n} and the right-hand panel of
Fig. \ref{fig:appendix} for the case of a Kerr black hole). This point
will be referred to as the ``slope point'' of the parametric curve
$R_{\psi}$ representing the shadow.

To compute the position of this point in the $(\alpha, \beta)$
coordinates we simply need to find the solution to the equation
\begin{equation}
\left.\frac{d\beta}{d\alpha}\right|_{\psi_{_S}} = 0\,,
\end{equation}
or, equivalently, solve for the differential equation
\begin{equation}
\label{slope_eq}
\frac{dR_{\psi}}{d\psi}\sin\psi + R_{\psi} \cos\psi = 0\,.
\end{equation}

Using the expansion in terms of Legendre polynomials
(\ref{leg_exp}), we can rewrite equation (\ref{slope_eq}) as
\begin{equation}
\sum_{\ell=0}^{\infty}c_\ell P_\ell (x) x
-\sum_{\ell=0}^{\infty}c_\ell \frac{dP_\ell (x)}{dx} (1-x^2)= 0
\label{poleq}\,,
\end{equation}
where we have set $x := \cos\psi$. The solutions of (\ref{poleq}) provide
the positions of all the possible slope points in the parametric curve,
and the solution is unique in the case in which the shadow $R(\psi)$ is
convex. The corresponding coordinates of the point $S$ are then
\begin{align}
R_{_S} =& \sum_{\ell=0}^\infty c_\ell P_\ell(x_{_S})\,,\\
\psi_{_A} =& \cos^{-1} (x_{_S})\,.
\end{align}

As for the second distortion parameter in Sect. \ref{dp_III}, we set $E$
to be the centre of the circle passing through the points $A$, $S$, and
$S'$, where $S'$ the point is symmetric to the point $S$ with respect to
the $\alpha$ axis. Using the condition $AE=ED$, we obtain the solution
\begin{align}
R_{_D} =&\left|\frac{\mathscr{A}^2 -(\sum_{\ell=0}^\infty c_\ell P_\ell (x_{_S}))^2}
{2 \sum_{\ell=0}^\infty c_\ell (1-P_\ell (x_{_S}) x_{_S})}\right|\,, \label{rrthree}\\
\psi_{_D} =& \cos^{-1} \left(\frac{R_{_D}}{|R_{_D}|}\right)\,.
\end{align}

Also for this third case, the radius of the circle $R_{s,_{\rm III}}$ passing
through the three points $A$, $S$, and $S'$, the distortion parameter
$d_{s,_{\rm III}}$, and its dimensionless version $\delta_{s,_{\rm III}}$, have
respectively the form

\begin{align}
\label{RS_III}
&\hspace{-0.1cm}R_{s,_{\rm III}}=\nonumber\\
&\frac{\mathscr{A}^2 -2 x_{_S}
\mathscr{A}(\sum_{\ell=0}^\infty c_\ell P_\ell
(x_{_S}))+(\sum_{\ell=0}^\infty c_\ell P_\ell (x_{_S}))^2}{2
\sum_{\ell=0}^\infty c_\ell (1-P_\ell (x_{_S}) x_{_S})}\,,
\\
&\hspace{-0.1cm}d_{s,_{\rm III}}=2 R_{s,_{\rm III}}- (R_{_A}+R_{_C})=
\left(\sum_{\ell=0}^\infty c_\ell P_\ell
  (x_{_S})\right) \nonumber\\&\times
 \frac{(\sum_{\ell=0}^\infty c_\ell P_\ell (x_{_S})- x_{_S}
  \sum_{\ell=1}^\infty c_{2\ell-1} - \mathscr{A}\,\mathscr{C})}
{\sum_{\ell=0}^\infty c_\ell (1-P_\ell (x_{_S}) x_{_S})} \label{bambidistt}\,,
\\
\label{deltadef3}
&\hspace{-0.1cm}\delta_{s,_{\rm III}}=\frac{d_{s,_{\rm III}}}{R_{s,_{\rm III}}}=
2\left(\sum_{\ell=0}^\infty c_\ell P_\ell
  (x_{_S})\right)\nonumber\\&
\times\frac{(\sum_{\ell=0}^\infty c_\ell P_\ell (x_{_S}) - x_{_S}
  \sum_{\ell=1}^\infty c_{2\ell-1} -
  \mathscr{A}\,\mathscr{C})}{\mathscr{A}^2 -
  2 x_{_S} \mathscr{A} \sum_{\ell=0}^\infty
  c_\ell P_\ell (x_{_S}) + (\sum_{\ell=0}^\infty c_\ell P_\ell
  (x_{_S}))^2}\,.
\end{align}

We note that this definition is similar to the one proposed by
\citet{Hioki09}, who measure the dimensionless distortion of the shadow
as
\begin{equation}
\label{delthioki}
\delta_{s,_{\rm HM}} := \frac{d_{s,_{\rm HM}}}{R_{s,_{\rm HM}}} \,,
\end{equation}
where
\begin{equation}
d_{s,_{\rm HM}} := R_{\psi}(\psi = \pi) - R_{s,_{\rm HM}}\,,
\end{equation}
and with $R_{s,_{\rm HM}}$ being the radius of the circle passing through
the points $A$, $S$, and $S'$. The most important difference with respect
to the definition of~\citet{Hioki09} is that we here express the parametric
curve in terms of the general Legendre expansion (\ref{leg_exp}),
while~\citet{Hioki09} assume the knowledge of $R_{\psi}$ at $\psi = \pi$.

Also in this case, expressions (\ref{RS_III})--(\ref{deltadef3}) are not
easy to handle analytically. However, in the simplest case in which the
expansion (\ref{leg_exp}) has only two nonvanishing terms, such that
$c_0\neq0\neq c_1$ and $c_\ell=0$ for $\ell \geq 2$, Eq. (\ref{poleq})
takes the more compact form
\begin{equation}
2 c_1 x^2 +c_0 x -c_1 = 0 \,,
\end{equation}
with solution
\begin{equation}
x_{_S}=-\frac{c_0}{4c_1}\pm\sqrt{\frac{c_0^2}{16 c_1^2}+\frac12}\,,
\end{equation}
and where the $+$ or $-$ signs refer to when $c_1>0$ and $c_1<0$,
respectively. The corresponding quantities $R_{s,_{\rm III}}$,
$d_{s,_{\rm III}}$ and $\delta_{s,_{\rm III}}$ have in this case the
following form
\begin{align}
\label{def3_comp_1}
R_{s,_{\rm III}} =& \frac{2 c_0^2+c_1^2(1+x_{_S})+2c_0 c_1 (1+x_{_S})}{2 [c_0+c_1
    (1+x_{_S})]} \,,\\
\label{def3_comp_2}
d_{s,_{\rm III}} =& \frac{c_1^2(1+x_{_S})}{c_0+c_1(1+x_{_S})}\,,\\
\label{def3_comp_3}
\delta_{s,_{\rm III}} =&\frac{2c_1^2(1+x_{_S})}{2
  c_0^2+c_1^2(1+x_{_S})+2c_0 c_1 (1+x_{_S})}= \delta_{1,_{\rm III}}\,.\ \
\end{align}

If the shadow is perfectly circular with radius $c_0$, then $c_1=0$ and
expressions (\ref{def3_comp_1})--(\ref{def3_comp_3}) show that
$R_{s,_{\rm III}}=c_0$, $d_{s,_{\rm III}}=0=\delta_{s,_{\rm III}}$, as
expected.

\section{Application of the formalism to black hole space-times}
\label{metriccc}

Having constructed a general formalism that allows us to describe in a
coordinate-independent manner the black hole shadow and to measure its
deformation, we are now ready to apply such a formalism to the specific
case of some well-known space-time metrics referring to axisymmetric
black holes. In particular, we will obviously start with the application
of the formalism to a rotating (Kerr) black hole (in
Sect. \ref{Kerr_BH}), to move over to a Bardeen black hole and to a
Kerr-Taub-NUT black hole in Sect. \ref{Bardeen_BH}. We note that we do
not consider these last two examples of black holes because they are
particularly realistic, but simply because they offer analytic line
elements on which our formalism can be applied.

\subsection{Kerr black hole}
\label{Kerr_BH}

We start with the Kerr space-time, whose line element in Boyer-Lindquist
coordinates reads
\begin{align}
\label{metricKerr}
ds^2 =& - \left(1 - \frac{2 M r}{\Sigma}\right) dt^2
- \frac{4 a M r \sin^2 \theta}{\Sigma} dt d\phi \nonumber\\
& + \left(r^2 + a^2 + \frac{2 a^2 M r \sin^2\theta}{\Sigma}\right)
\sin^2\theta d\phi^2 \nonumber\\
&
+ \frac{\Sigma}{\Delta} dr^2 + \Sigma d\theta^2\,,
\end{align}
where
\begin{equation}
\label{metricfunctions}
\Sigma := r^2 + a^2 \cos^2\theta \,, \qquad
\Delta := r^2 - 2 M r + a^2 \,,
\end{equation}
with $M$ being the mass of the black hole and $a := J/M$ its specific
angular momentum.

Since the shape of the shadow is ultimately determined by the innermost
unstable orbits of photons, hereafter we will concentrate on their
equations for photons. In such a space-time, the corresponding geodesic
equations take the form
\begin{align}
\Sigma \bigg(\frac{d t}{d \lambda}\bigg)
=& \frac{A E - 2 a M r L_z}{\Delta} \,, \\
\Sigma^2\bigg(\frac{d r}{d \lambda}\bigg)^2
=& \mathcal{R} \,,   \label{eq-radial} \\
\Sigma^2\bigg(\frac{d \theta}{d \lambda}\bigg)^2
=& \Theta \,, \\
\Sigma \bigg(\frac{d \phi}{d \lambda}\bigg)
=& \frac{2 a M r E}{\Delta}+
\frac{\left(\Sigma - 2 M r\right) L_z  }{\Delta\sin^2\theta}  ,
\end{align}
where $\lambda$ is an affine parameter,
\begin{align}
\label{eq-R}
\mathcal{R} :=& E^2 r^4+\left(a^2E^2-L_z^2-\mathcal{Q}\right)r^2
\nonumber\\
& +2 M \left[(aE-L_z)^2+\mathcal{Q}\right]r-a^2\mathcal{Q}\,, \\
\Theta :=& \mathcal{Q} \left(a^2 E^2 - L_z^2 \csc^2 \theta \right)
\cos^2 \theta \,, \\
A :=& \left(r^2 + a^2\right)^2 - a^2 \Delta \sin^2 \theta \,,
\end{align}
and $E$ and $L_z$ are the photon's energy and the angular momentum,
respectively. The quantity $\mathcal{Q}$
\begin{align}
\mathcal{Q} = p_\theta^2+\cos^2\theta
\bigg(\frac{L_z^2}{\sin^2\theta}-a^2E^2\bigg) \,
\end{align}
is the so-called Carter constant and $p_\theta := \Sigma
{d\theta}/{d\lambda}$ is the canonical momentum conjugate to $\theta$.

Using these definitions, it is possible to determine the unstable orbits
as those satisfying the conditions 
\begin{equation}
\label{condition}
\mathcal{R}(\bar{r}) = \frac{\partial \mathcal{R}(\bar{r})}{\partial r}=0
\,, \quad {\rm and} \quad \frac{\partial^2 \mathcal{R}(\bar{r})}{\partial
r^2} \geq 0 \,,
\end{equation}
where $\bar{r}$ is the radial coordinate of the unstable orbit. Introducing
now the new parameters $\xi := L_z/E$ and $\eta := \mathcal{Q}/E^2$, the
celestial coordinates $\alpha$ and $\beta$ of the image plane of the
distant observer are given by \citep{Bardeen72}
\begin{equation}
\alpha = \frac{\xi}{\sin i}\,,  \quad
\beta = \pm (\eta+a^2\cos^2 i-\xi^2\cot^2 i)^{1/2}\,,
\end{equation}
where $i$ is the inclination angle of the observer's plane, that is, the
angle between the normal to the observer's plane and the black hole's
axis of rotation. In the case of the Kerr space-time (\ref{metricKerr})
and after using the conditions (\ref{condition}), one can easily find
that the values of $\xi$ and $\eta$ relative to the circular orbit ($c$)
are \citep{Young76,Chandrasekhar98}
\begin{align}
\label{xieta}
\xi_c =& \frac{M(\bar{r}^2-a^2)-\bar{r}(\bar{r}^2-
2M\bar{r}+a^2)}{a(\bar{r}-M)} \,, \\
\eta_c =& \frac{\bar{r}^3 [4a^2M-\bar{r}(\bar{r}-3M)^2]}
{a^2(\bar{r}-M)^2} \,,
\end{align}

\begin{figure*}
\begin{center}
  \includegraphics[width=0.8\columnwidth]{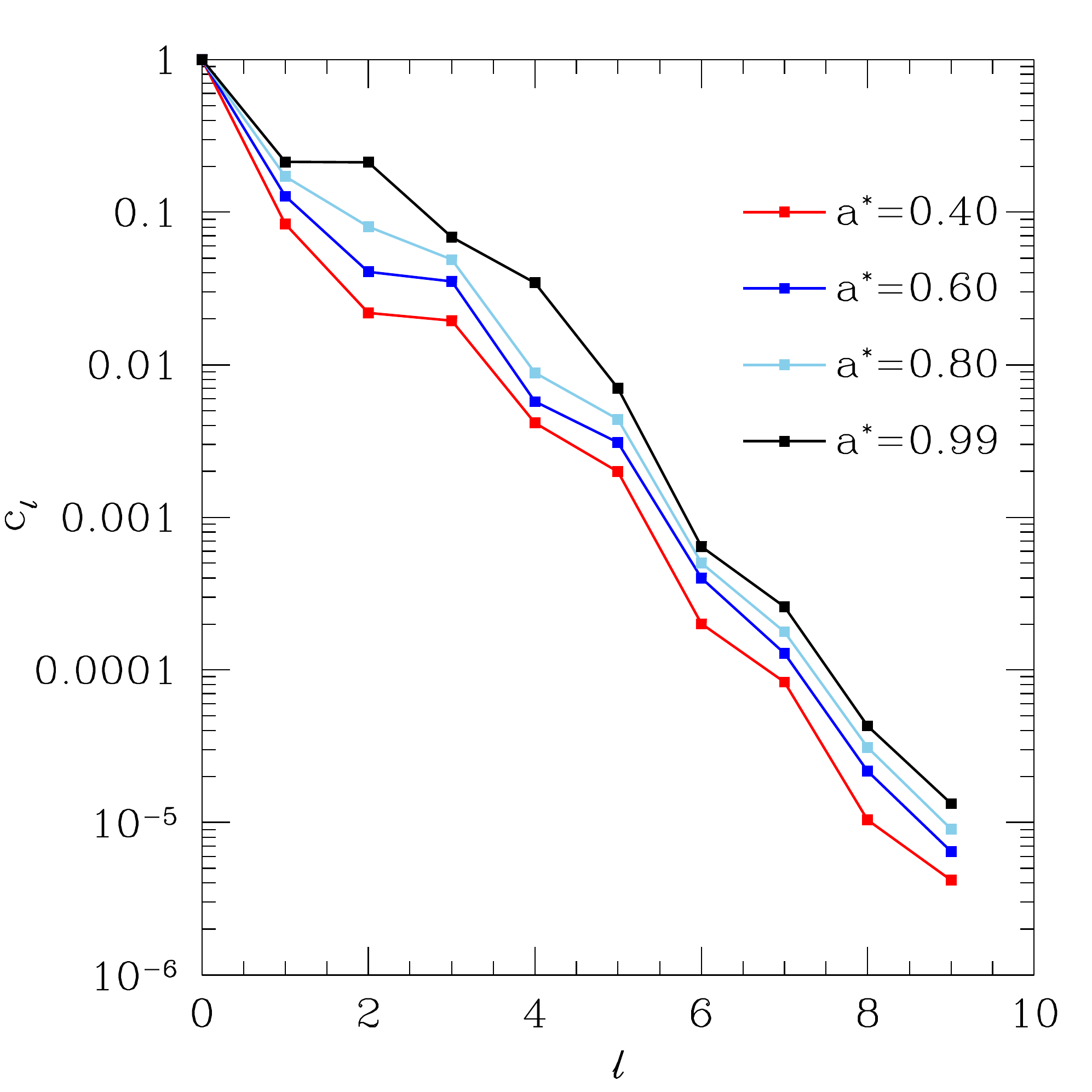}
  \hskip 2.0cm
  \includegraphics[width=0.8\columnwidth]{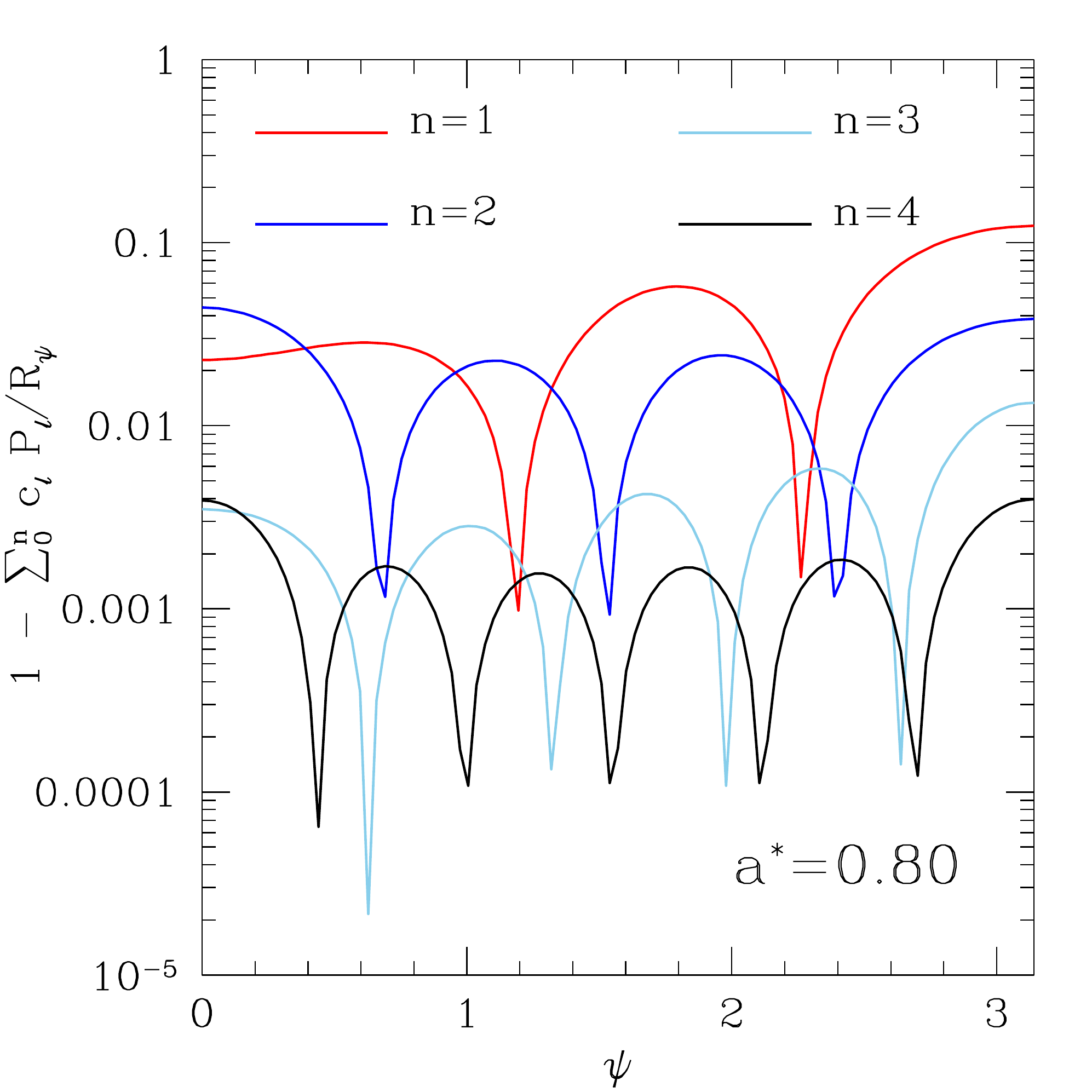}
\end{center}
\caption{\textit{Left-hand panel:} Magnitude of the expansion coefficients
  $c_\ell$ of the polar curve and shown as a function of the expansion
  order. Note the very rapid (exponential) convergence of the
  expansion. The coefficients are computed for a Kerr black hole and
  different lines refer to different values of the spin parameter $a^*$.
  \textit{Right-hand panel:} Relative differences between the polar curve for
  the black hole shadow as constructed from expressions
  (\ref{ruch})--(\ref{psiuch}) and the corresponding curve obtained from
  the expansion, \ie $1-\sum^n_0 c_{\ell} P_{\ell}/R_{\psi}$. Different
  lines refer to different truncations of the expansion and show that
  three coefficients are sufficient to obtain deviations of a few
  percent.}
\label{fig4}
\end{figure*}

Next, to investigate the shape of the black hole shadow we introduce the
generic celestial polar coordinates $(R', \psi')$ (\cf
Sect. \ref{general:form} and Fig. \ref{cartoon_OO'}) defined as
\begin{align}
R' =& (\alpha^2+\beta^2)^{1/2}\,, \\
\psi' =& \tan^{-1}\left(\frac{\beta}{\alpha}\right)\,.
\end{align}
Assuming for simplicity that the observer is in the equatorial plane, \ie
that $i=\pi/2$, then in terms of the $(R',\psi')$ coordinates the shadow
of black hole can be described as (hereafter we will set $M=1$)
\begin{align}
\label{ruch}
R' =& \frac{(2 r^4+2 a^2 r-6 r^2+a^2+a^2 r^2)^{1/2}}{r-1}\,,\\
\label{psiuch}
\psi'=& \tan^{-1}\left(\frac{\{r^3 [4 a^2-r (r-3)^2]\}^{1/2}}
{r^2-a^2-r (r^2-2 r+a^2)}\right)\,.
\end{align}

\begin{figure*}
\begin{center}
\includegraphics[width=0.32\textwidth]{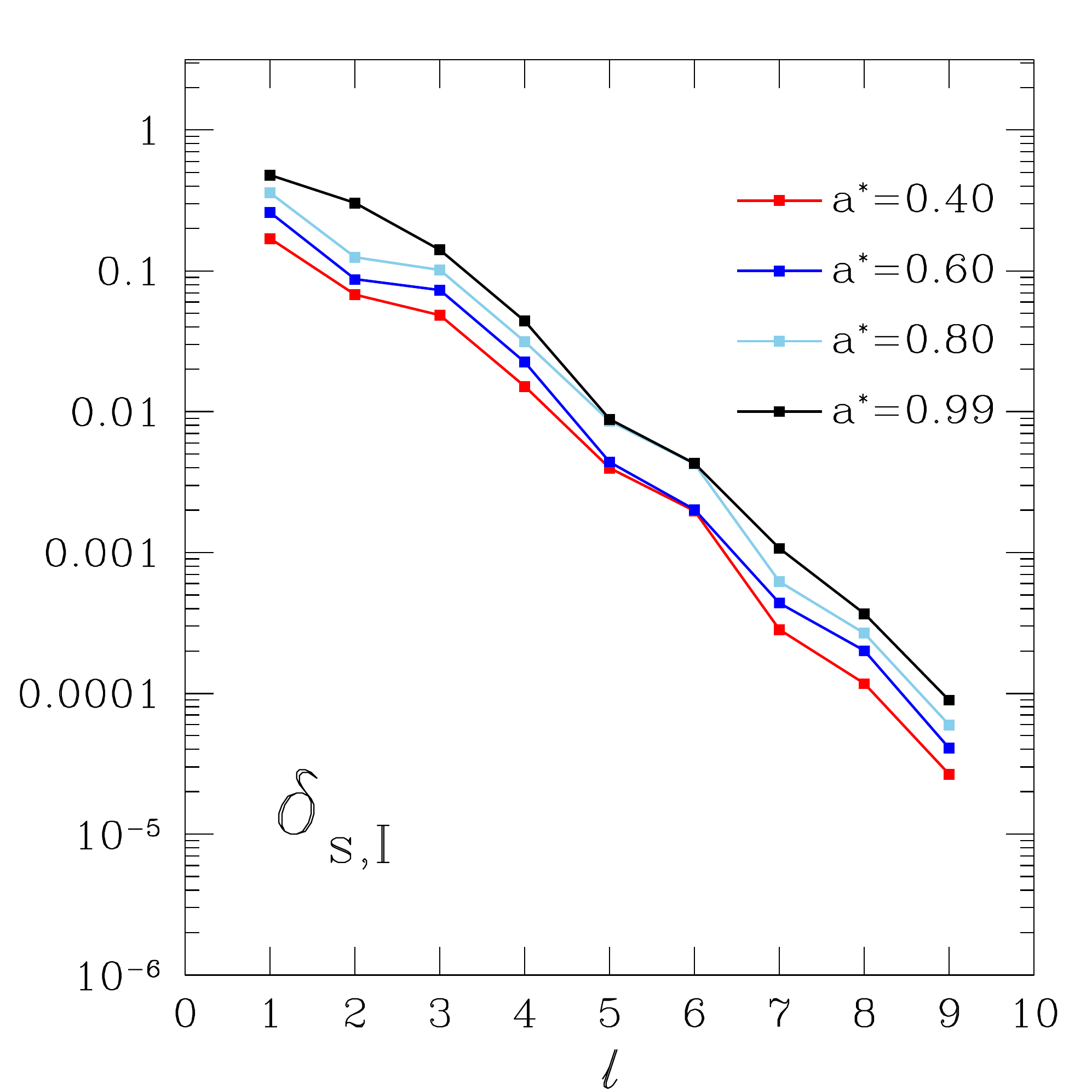}
\hskip 0.25cm
\includegraphics[width=0.32\textwidth]{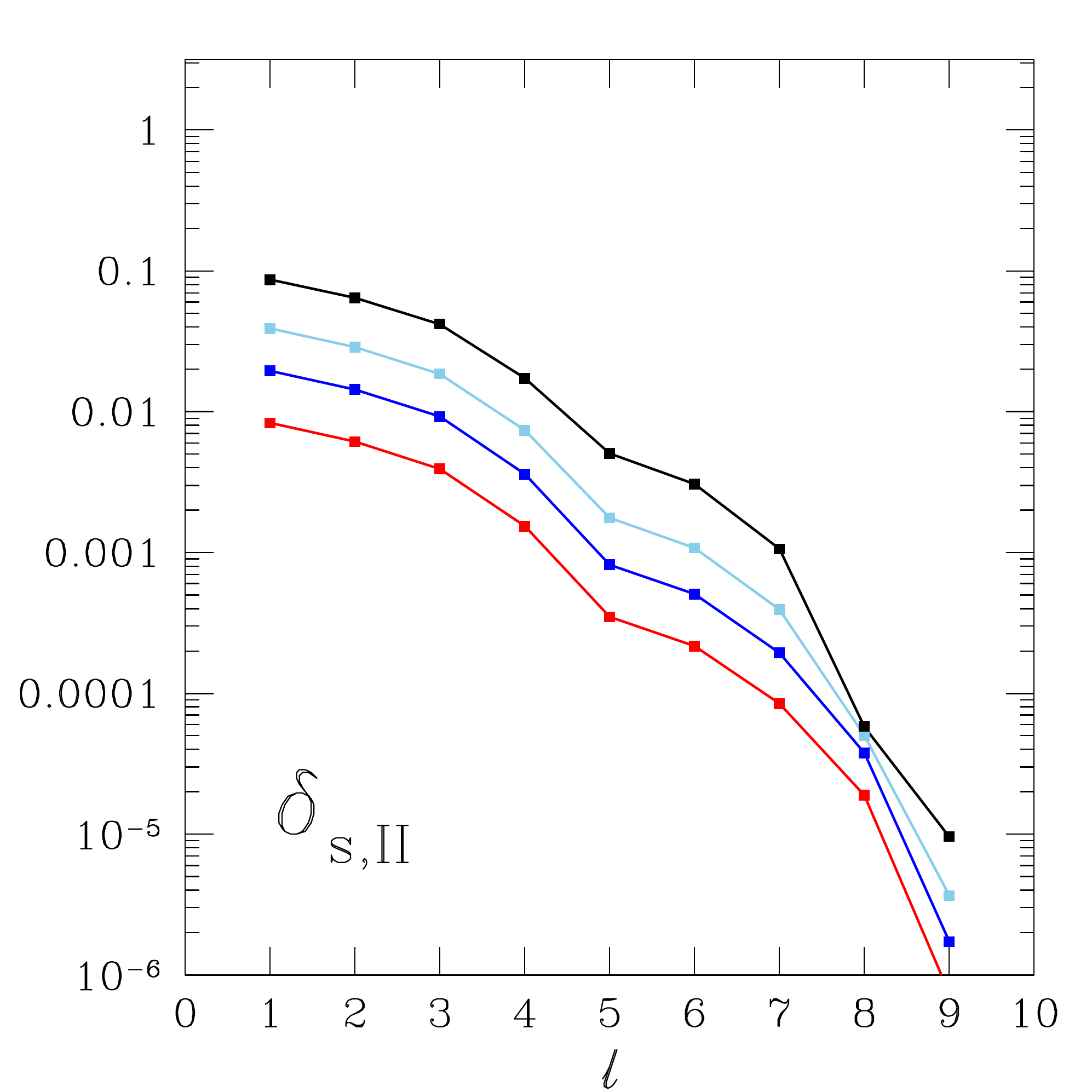}
\hskip 0.25cm
\includegraphics[width=0.32\textwidth]{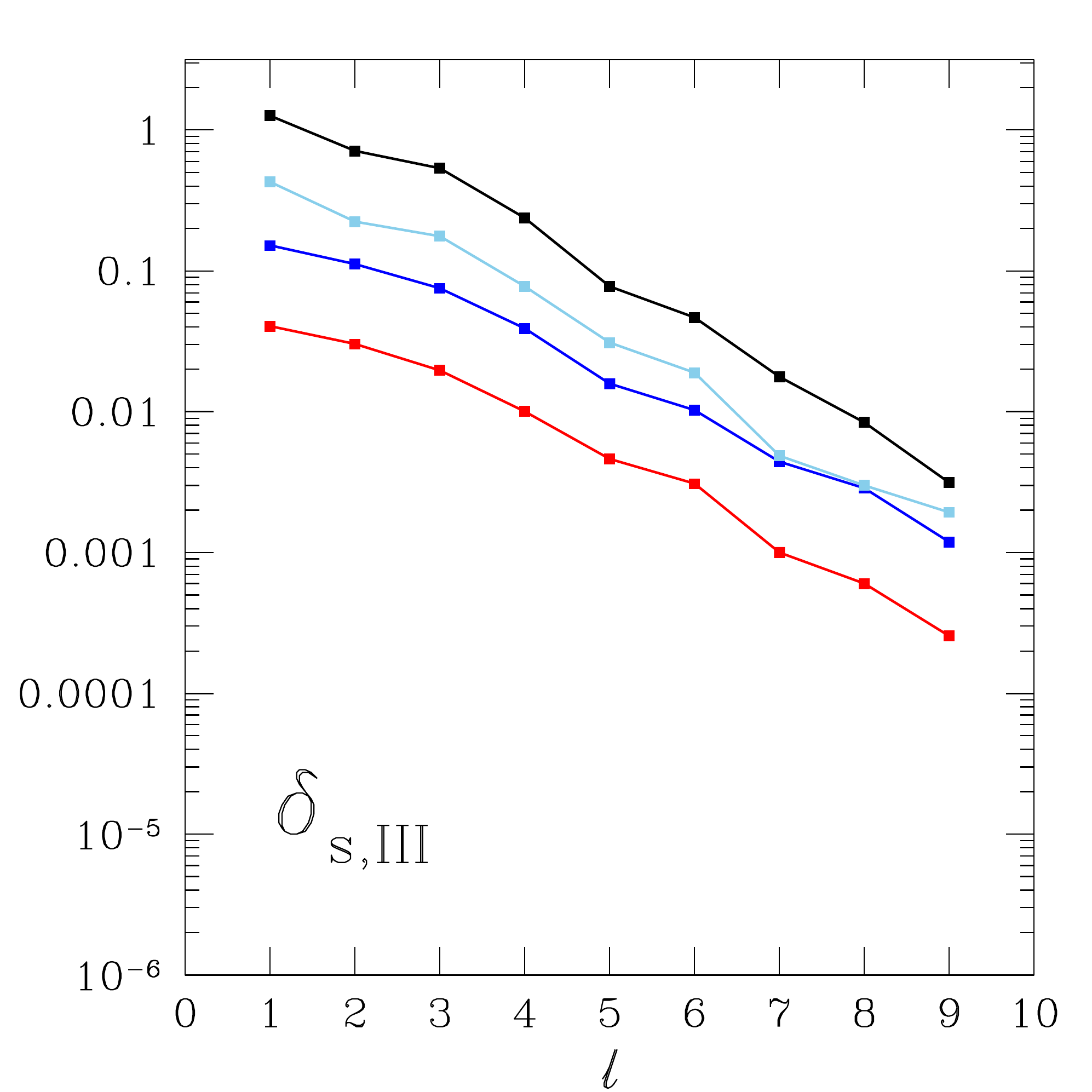}
\end{center}
\caption{Magnitude as a function of the expansion order of the three
  different distortion parameters $\delta_{s,_{\rm I}}$--$\delta_{s,_{\rm
      III}}$ defined as in Eqs. (\ref{deltadef1}), (\ref{deltadef2}), and
  (\ref{deltadef3}), thus measuring the difference between the Legendre
  expanded polar curves $R_{\psi,_{\rm I}}$--$R_{\psi,_{\rm III}}$ and
  the reference circles of radii $R_{s,_{\rm I}}$--$R_{s,_{\rm
      III}}$. All curves refer to a Kerr black hole and different colours
  are used to represent different values of the spin parameter.}
 \label{dist_all_defs}
\end{figure*}

Making use of the procedure described in Sect.~\ref{general:form}, it is
straightforward to determine the coordinates
(\ref{rzero})--(\ref{psizero}) of the effective centre of the shadow, and
to perform the Legendre expansion (\ref{leg_exp}). To the best of
our knowledge, no analytic expression exists to cast the coordinates
(\ref{ruch})--(\ref{psiuch}) as a polar curve
$R_{\psi}=R(\psi)$. However, such a curve can be easily constructed
numerically and from it the Legendre expansion (\ref{leg_exp}) can be
computed.

Figure~\ref{fig4} summarizes the results of our approach by reporting in
the left-hand panel and in a logarithmic scale the normalized values of the
expansion coefficients $c_\ell$ as a function of the Legendre order
$\ell$. Different curves refer to the different values considered for the
dimensionless spin parameter $a^* := J/M^2 = a/M$, which ranges from
$a^*=0.4$ (blue solid line) to $a^*=0.99$ (red solid
line). Interestingly, the series converges very rapidly (essentially
exponentially) and already with $\ell =4$, the contribution of
higher-order terms is of the order of $10^{-2}$, decreasing further to
$\sim 10^{-3}$ for $\ell=6$. Furthermore, even when considering the more
severe test of $a^*=0.99$, the expansion coefficient with $\ell=6$ is
only a factor 2-3 larger than the corresponding coefficient for a
slowly rotating black hole. The right-hand panel of Fig.~\ref{fig4} shows a
direct measure of the relative differences between the polar curve for
the black hole shadow as constructed from expressions
(\ref{ruch})--(\ref{psiuch}) and the corresponding curve obtained from
the expansion, \ie $1-\sum^n_0 c_{\ell} P_{\ell}/R_{\psi}$. Remarkably,
already when considering the first three terms in the expansion, \ie
$c_0, c_1$, and $c_2$, the relative difference is of a few percent (blue
line), and this further reduces to $10^{-3}$ when the expansion is
truncated at $n=4$ (black line).

\begin{figure*}
\begin{center}
  \includegraphics[width=0.8\columnwidth]{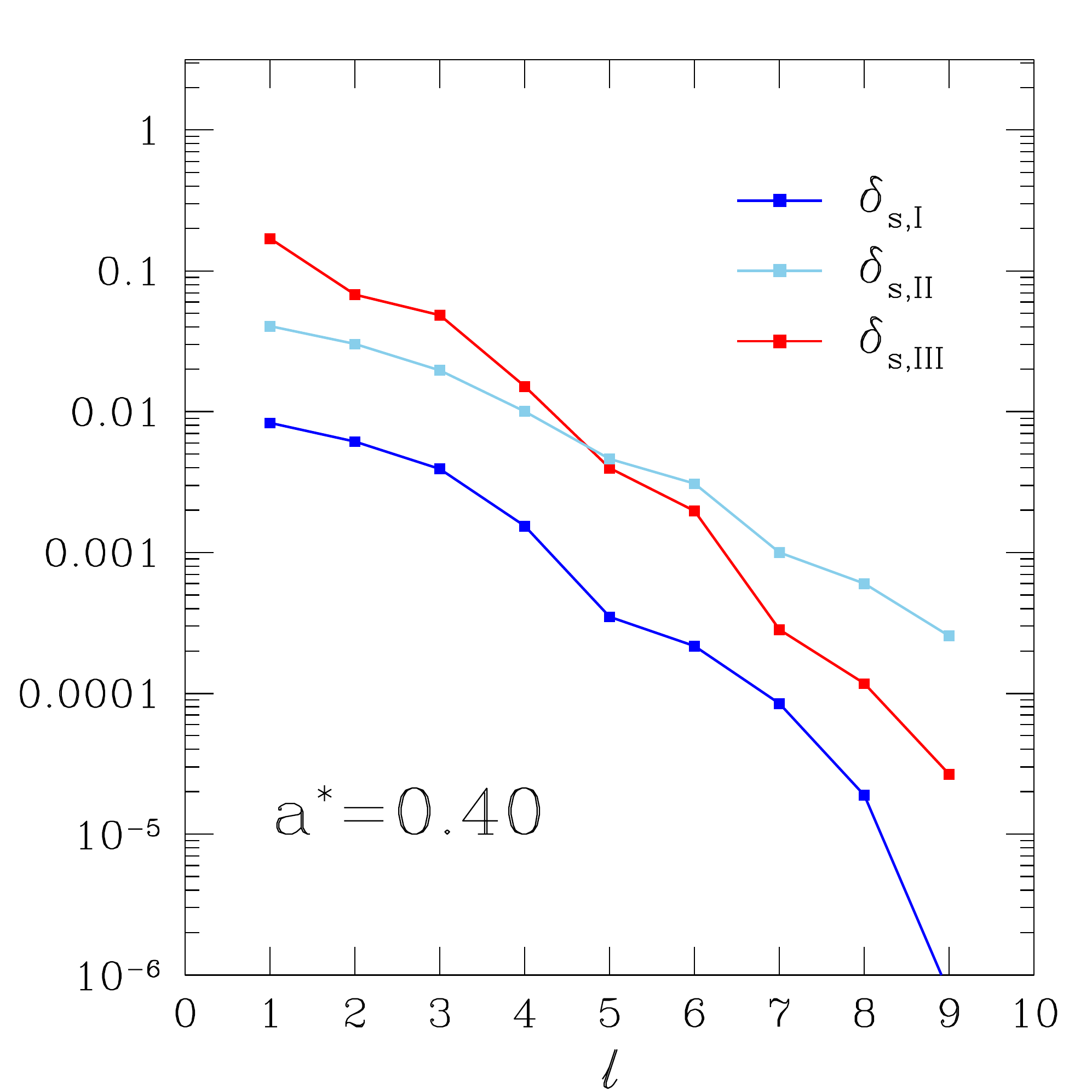}
  \hskip 2.0cm
  \includegraphics[width=0.8\columnwidth]{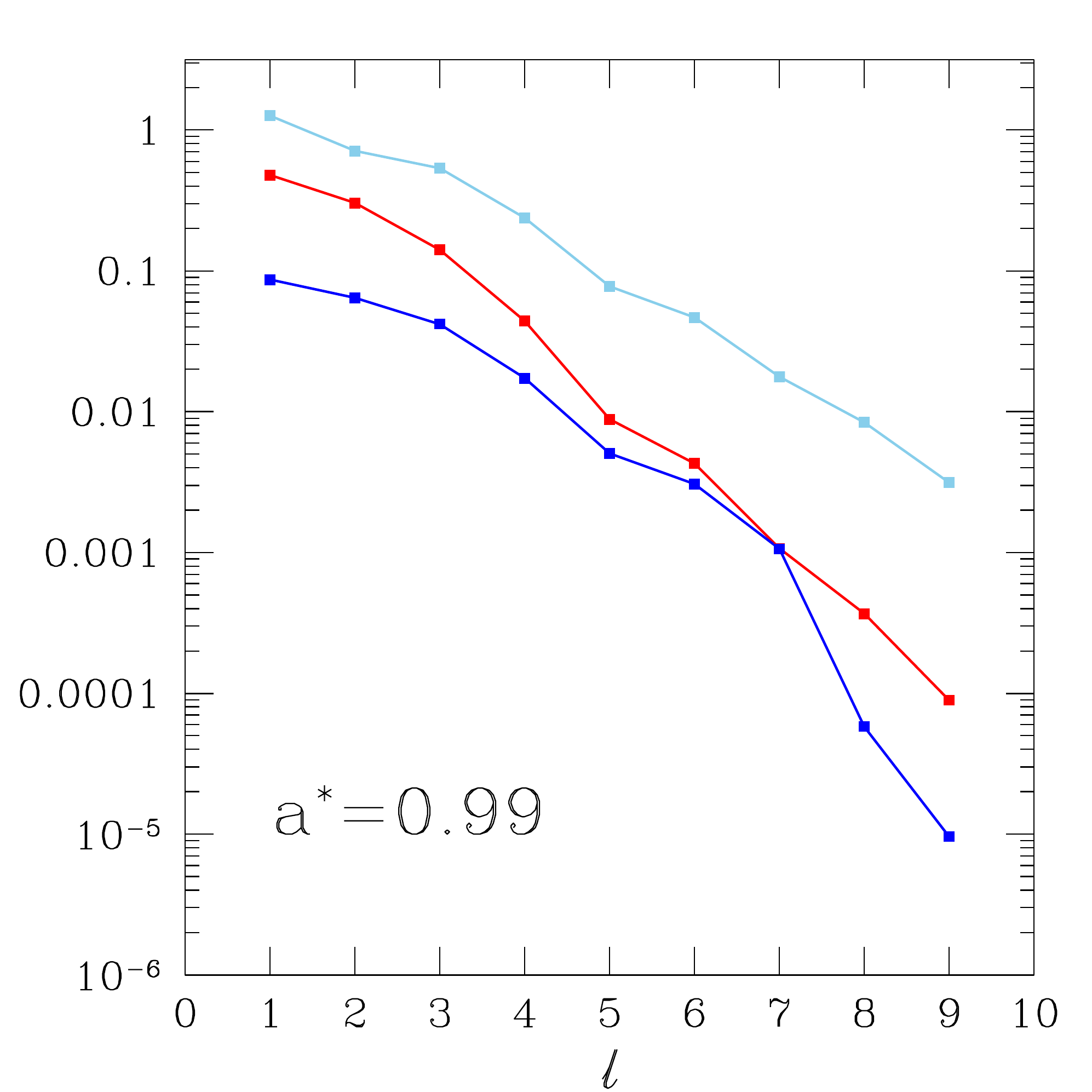}
\end{center}
\caption{Comparative view of the different distortion parameters
  $\delta_{s,_{\rm I}}$ (red line), $\delta_{s,_{\rm II}}$ (blue line),
  and $\delta_{s,_{\rm III}}$ (light-blue line). The left-hand and right-hand 
  panels show the values of the distortion parameters as a function of
  the expansion order (\cf Fig. \ref{dist_all_defs}), and refer to a Kerr
  black hole with spin $a^*=0.40$ and $a^*=0.99$, respectively.}
\label{dist_allinone}
\end{figure*}

In summary, Fig.~\ref{fig4} demonstrates that when considering a Kerr
black hole, the approach proposed here provides a coordinate independent
and accurate representation of the black hole shadow and that a handful
of coefficients is sufficient for most practical purposes. In the
following Sections we will show that this is the case also for other
axisymmetric black holes.

Before doing that, we show in Fig. \ref{dist_all_defs} the values of the
dimensionless distortion parameters as computed for the shadow of a Kerr
black hole and for increasing values of the expansion index $\ell$. The
three different panels are relative respectively to the parameters
(\ref{deltadef1}), (\ref{deltadef2}), and (\ref{deltadef3}), with the
different curves referring to values of the spin parameter $a^*$, ranging
from $a^*=0.4$ (blue solid line) to $a^*=0.99$ (red solid line). As one
would expect, for all values of $a^*$, each of the three distortion
parameters decreases as the expansion includes higher-order terms. At the
same time, because larger rotation rates introduce larger distortions in
the shadow, they also lead to larger values of the distortion parameters
for a fixed value of $\ell$.

Finally, Fig. \ref{dist_allinone} offers a comparative view of the
different distortion parameters for specific values of the spin
parameter, with the left-hand and right-hand panels referring to $a^*=0.4$ and
$a^*=0.99$, respectively. This view is rather instructive as it shows that
the different definitions lead to significantly different values of the
distortion, despite they all refer to the same parametric polar
curve. Furthermore, it helps appreciate that the distortion parameter
$\delta_{s,_{\rm II}}$ is systematically smaller than the other two and
hence not the optimal one. This is because a larger value of the
distortion parameter will increase the possibility of capturing the
complex structure of the shadow. The fact that the curves for
$\delta_{s,_{\rm I}}$ and $\delta_{s,_{\rm III}}$ intersect for the Kerr
black hole considered at $\ell=5$ implies that both distortion parameters
(\ref{deltadef1}) and (\ref{deltadef3}) are useful indicators, with
$\delta_{s,_{\rm III}}$ being the recommended choice for expansions
having $\ell \geq 5$.

\begin{figure*}
\begin{center}
  \includegraphics[width=0.8\columnwidth]{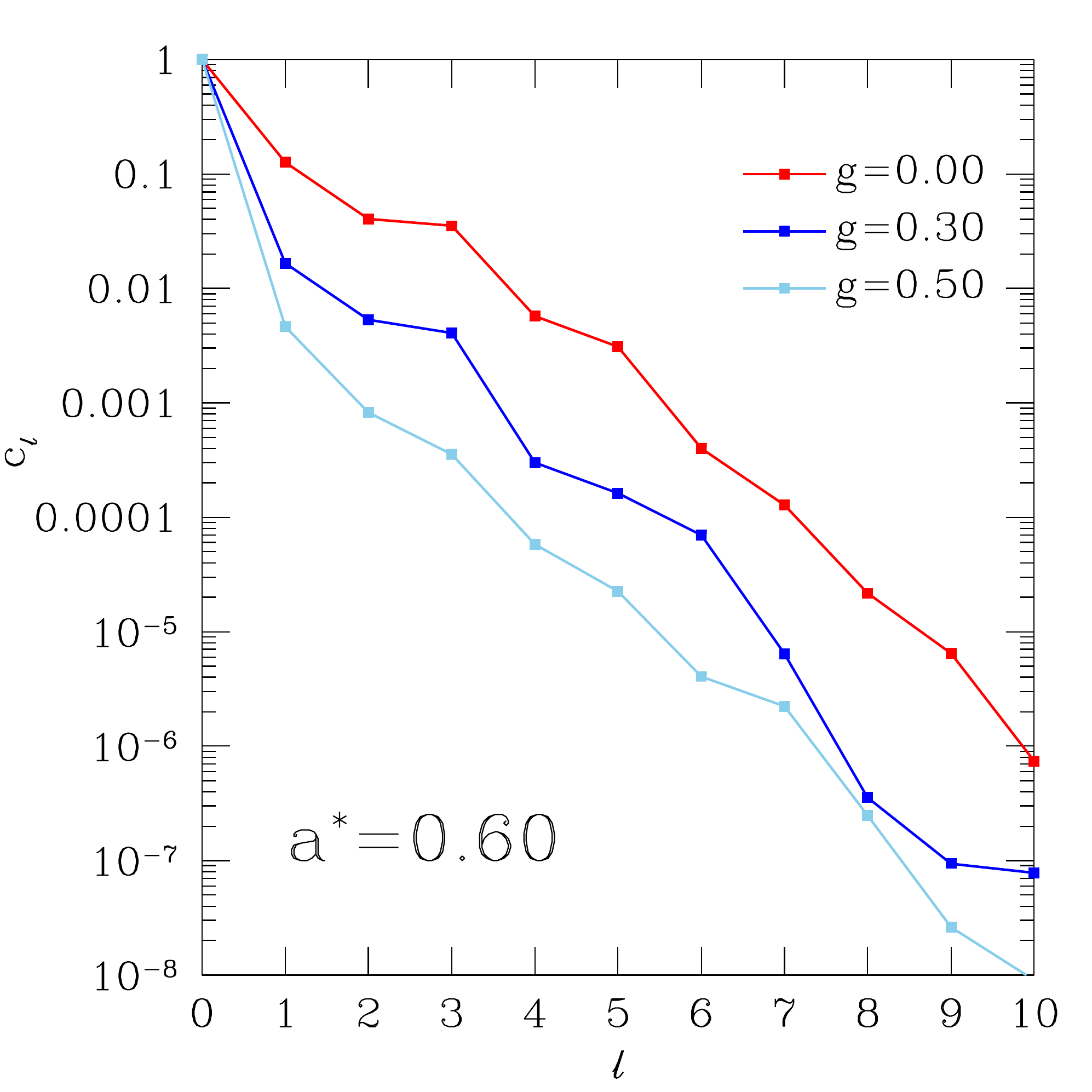}
  \hskip 2.0cm
  \includegraphics[width=0.8\columnwidth]{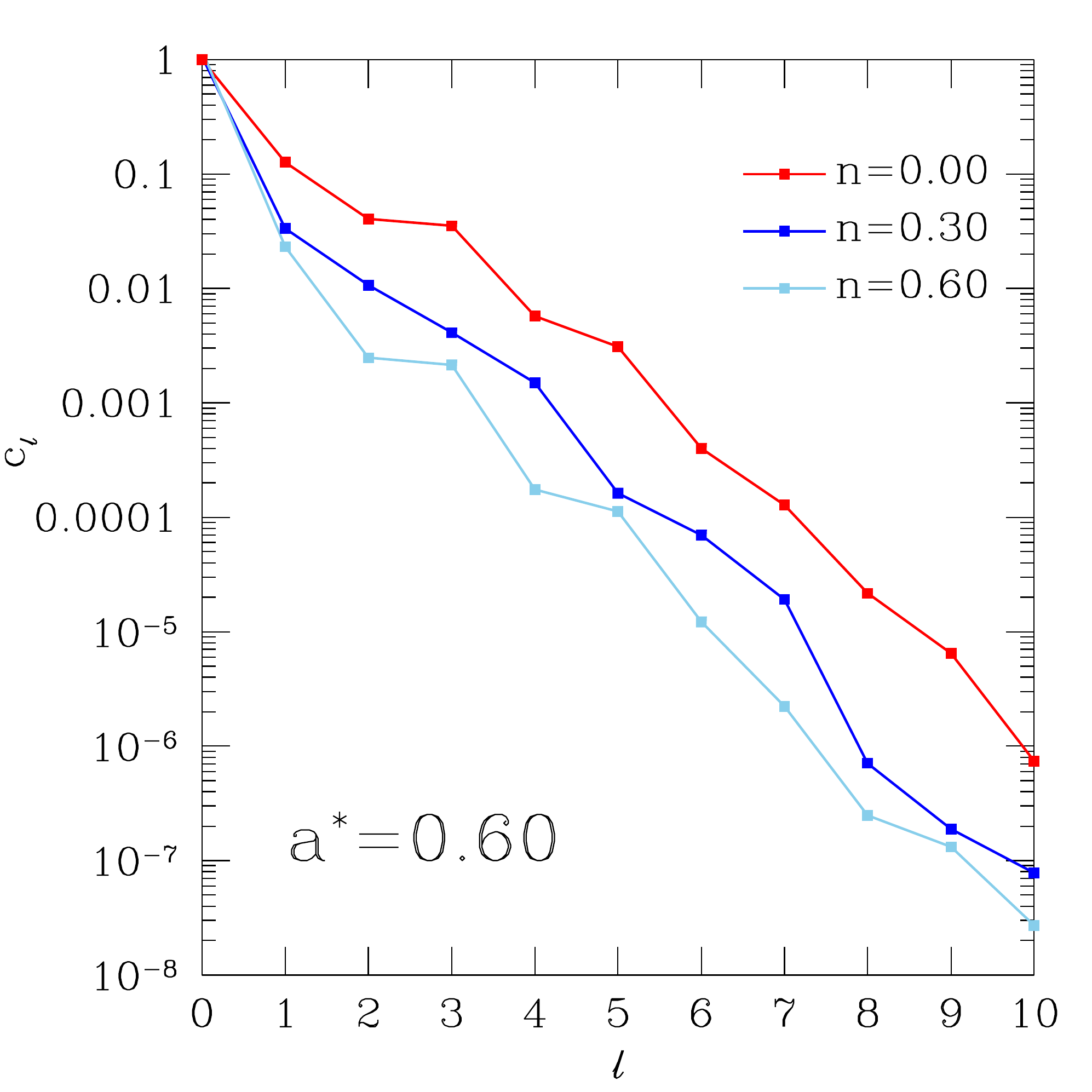}
\end{center}
\caption{\textit{Left-hand panel:} Magnitude of the expansion coefficients
  $c_\ell$ as a function of the expansion order $\ell$ for the different
  values of the magnetic charge of a Bardeen black hole: $g=0.00$ (red
  line), $g=0.30$ (blue line), and $g=0.50$ (light-blue line). All lines
  refer to a fixed value of the rotation parameter $a^*=0.60$ (\cf left
  panel of Fig. \ref{fig4}). \textit{Right-hand panel:} The same as in the
  left-hand panel but for a Kerr-Taub-NUT black hole. Different curves refer
  to the different values of the NUT parameter: $n=0.00$ (red line),
  $n=0.30$ (blue line), and $n=0.60$ (light-blue line). All lines refer
  to a fixed value of the rotation parameter $a^*=0.60$.}
\label{cells_B_NUT}
\end{figure*}

\subsection{Bardeen and Kerr-Taub-NUT black holes}
\label{Bardeen_BH}

We continue our application of the formalism developed in
Sects. \ref{general:form} and \ref{dist_params} by considering the
space-time of a rotating Bardeen black hole \citep{Bardeen68}. We recall
that in Boyer-Lindquist coordinates, the metric of a Kerr and of a
Bardeen black hole differ uniquely in the mass, which needs to be
modified as~\citep{Bambi2013,Tsukamoto14}
\begin{equation}
M\rightarrow m=M\left(\frac{r^2}{r^2+g^2}\right)^{3/2}\,,
\end{equation}
where the parameter $g$ is the magnetic charge of the nonlinear
electrodynamic field responsible for the deviation away from the Kerr
space-time.

The impact parameters $\xi$ and $\eta$ relative to the circular orbit are
in this case~\citep{Tsukamoto14}
\begin{align}
\label{xic_bardeen}
\xi_c =& \frac{m [(2-f) \bar{r}^2-f a^2]-\bar{r}(\bar{r}^2-2m
\bar{r}+a^2)}{a(\bar{r}-fm)} \,, \\
\label{etac_bardeen}
\eta_c =& \frac{\bar{r}^3 \{4(2-f)a^2
m-\bar{r}[\bar{r}-(4-f)m]^2\}}{a^2(\bar{r}-fm)^2}\,,
\end{align}
and can be taken to define the shadow of black hole. Note that $m$ and
$f$ are functions of the unstable circular radius $\bar{r}$ and are given
by
\begin{align}
m =& m(\bar{r})=M \left(\frac{\bar{r}^2}{\bar{r}^2+g^2}\right)^{3/2}\,,\\
f =& f(\bar{r})=\frac{\bar{r}^2+4 g^2}{\bar{r}^2+g^2}\,.
\end{align}

In complete analogy, we can consider a Kerr-Taub-NUT black hole with
nonvanishing gravitomagnetic charge $n$ and specific angular momentum
$a:=J/M$. The corresponding metric is given by~\citep{Newman63}
\begin{align}
ds^2 = &  -\frac{1}{\Sigma} \left(\Delta - a^2 \sin^2\theta
\right)dt^2 + \Sigma \left(\frac{dr^2}{\Delta} +
d\theta^{2}\right)\nonumber \\
&
+ \frac{1}{\Sigma}\left[ (\Sigma+a\chi)^2 \sin^2\theta  -
\chi^2\Delta \right]d\phi^2\nonumber \\
& +\frac{2}{\Sigma}(\Delta\chi-a(\Sigma+a\chi)\sin^2\theta)dt
d\phi , \label{metric}
\end{align}
where the functions $\Delta, \Sigma,$ and $\chi$ are now defined as
\begin{align}
\Delta :=& r^2 + a^2  - n^2 - 2 M r\,, \nonumber \\
\Sigma :=&r^2+(n+a \cos\theta)^2 \,, \nonumber \\
\chi :=& a \sin^2\theta- 2 n \cos\theta\,.
\end{align}

In this case, the impact parameters $\xi$ and $\eta$ for the circular
orbits are given by~\citep{Abdujabbarov2013} 
\begin{align}
\xi_c =& \frac{a^{2}(1+\bar{r})+\bar{r}^{2}(\bar{r}-3)+n^{2}(1-3
  \bar{r}) }{a(1-\bar{r})}\,,
\label{eqxi} \\
\eta_c =& \frac{1}{a^{2}(\bar{r}-1)}\bigg\{\bar{r}^{3}[4a^{2}-
  \bar{r}(\bar{r}-3)^{2}]-n^{2}\big[4 \bar{r}^2 a^2 \nonumber \\ 
&+(1-3 \bar{r})(n^2(1-3 \bar{r})-6 \bar{r}^2+4 a^2 \bar{r}+2
  \bar{r}^3)\big]\bigg\} \,,
\label{eqetata}
\end{align}
and define the shadow of the Kerr-Taub-NUT black hole.

Applying the formalism described in Sect.~\ref{general:form}, it is
possible to compute the coefficients of the Legendre expansion
(\ref{leg_exp}) also for a Bardeen and for a Kerr-Taub-NUT black
hole. The numerical values of these coefficients are reported as a
function of the expansion order $\ell$ in Fig. \ref{cells_B_NUT}, where
the left-hand panel refers to a Bardeen black hole, while the right-hand one to a
Kerr-Taub-NUT black hole. More specifically, the different lines in the
left-hand panel refer to different values of the magnetic charge: $g=0.00$
(red line), $g=0.30$ (blue line), and $g=0.60$ (light-blue line); all
lines refer to a fixed value of the rotation parameter $a^*=0.60$. Very
similar is also the content of the right-hand panel of Fig. \ref{cells_B_NUT},
which is however relative to a Kerr-Taub-NUT black hole. The different
curves in this case refer to the different values of the NUT parameter:
$n=0.00$ (red line), $n=0.30$ (blue line), and $n=0.60$ (light-blue
line); once again, all lines refer to a fixed value of the rotation
parameter $a^*=0.60$.

In analogy with what shown in the left-hand panel of Fig.~\ref{fig4} for a
Kerr black hole, also for these black holes the series converges very
rapidly and already with $\ell =4$, the contribution of higher-order
terms is of the order of $10^{-3}$, decreasing further to $\sim 10^{-5}$
for $\ell=6$, even when considering higher larger magnetic charges or NUT
parameters. Furthermore, in analogy with the right-hand panel of
Fig.~\ref{fig4}, we have checked that the relative differences between
the polar curves for the shadow constructed from expressions
(\ref{xic_bardeen})--(\ref{etac_bardeen}) and
(\ref{eqxi})--(\ref{eqetata}), and the corresponding curve obtained from
the expansion, \ie $1-\sum^n_0 c_{\ell} P_{\ell}/R_{\psi}$, is below
$10^{-2}$ when $n=2$ (not shown in Fig.~\ref{cells_B_NUT}); this
difference further reduces to $10^{-4}$ when the expansion is truncated
at $n=4$.

In conclusion, also Fig.~\ref{cells_B_NUT} demonstrates that the approach
proposed here provides a coordinate independent and accurate
representation of black hole shadows in space-times other than the Kerr
one.


\section{Comparison with noisy observational data}
\label{comparison}

All of the considerations made so far have relied on the assumption that
the shadow is a well-defined one-dimensional curve (\cf discussion in
Sect \ref{general:form}). In practice, however, this is not going to be
the case as the astronomical observations will have intrinsic
uncertainties that, at least initially, will be rather large. It is
therefore natural to ask how the formalism presented here will cope with
such uncertainties. More precisely, it is natural to ask whether it will
still be possible to determine the effective centre of a noisy polar
curve and then determine from there its properties. Although a very
obvious and realistic problem, this concern is systematically ignored in
the literature, where the shadow is traditionally assumed to have no
uncertainty due to the observational measurements.

While awaiting for actual observational data, we can straightforwardly
address this issue in our formalism and mimic the noisiness in the
observational data by considering the polar curve as given by the
Legendre expansion (\ref{leg_exp}), where however the different
coefficients $c_{\ell}$ are artificially perturbed. More specifically, we
express the shadow via the new expansion
\begin{equation}
\label{perturbed}
R_{\psi}= \sum_{\ell=0}^{\infty} c_\ell (1+\Delta_c)
P_{\ell}(\cos\psi)\,,
\end{equation}
where $\Delta_c$ is a random real number chosen uniformly in the range
$[-\Delta_{\rm max}, \Delta_{\rm max}]$. In this way, our putative polar
curve representing the shadow will be distorted following a random
distribution and we have optimistically assumed a variance of only 5\%,
\ie $\Delta_{\rm max}=0.05$. Of course, there is no reason to expect that
the error distribution in the actual observational data will be uniform,
but assuming a white noise is for us the simplest and less arbitrary
choice.

\begin{figure}
\begin{center}
\includegraphics[width=0.9\columnwidth]{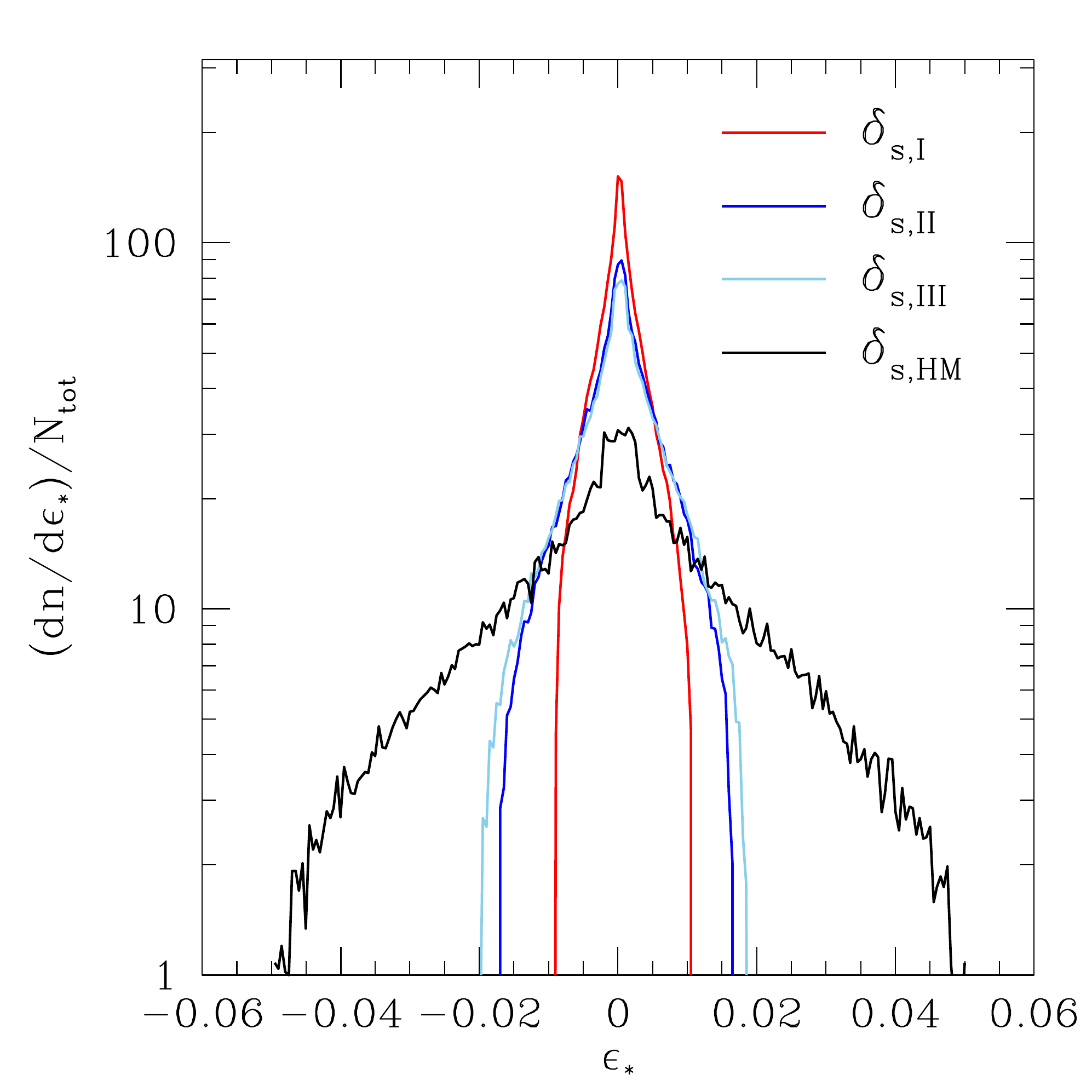}
\caption{Comparison of probability density distributions of the errors
  $\epsilon_*$ computed in the measurement of the distortion parameters
  $\delta_{s,_{\rm I}}-\delta_{s,_{\rm III}}$ for a Kerr black hole
  shadow reconstructed using a perturbed expansion [\cf
    Eq. (\ref{perturbed})]. Also shown is the distribution of the error
  measured when using the distortion parameter introduced by
  \citet{Hioki09} and that has a larger variance.}
\label{distribution}
\end{center}
\end{figure}

With the setup described above and the formalism discussed in the
previous Sections, we have considered a reference shadow of a Kerr black
hole with spin parameter $a/M=0.99$ and have reproduced it after
truncating the expansion (\ref{perturbed}) at $\ell=9$, which is more
than sufficient given the accuracy obtained at this order (\cf
Fig. \ref{fig4}). We have therefore constructed a very large number of
such realizations of the observational shadow after making use of $N_{\rm
  tot}=10^5$ draws of the random deformation $\Delta_c$. For each
putative observed reconstructed shadow we have computed the distortion
parameters $\delta_{s,_{\rm I}}-\delta_{s,_{\rm III}}$ defined in
Eqs. (\ref{deltadef1}), (\ref{deltadef2}), and (\ref{deltadef3}), as well
as the distortion definition of \citet{Hioki09} and defined in
Eq. (\ref{delthioki}).

For each of the shadow realizations we have therefore computed the
measurement error as
\begin{equation}
\epsilon_* := \delta_s - \delta_{s,*}\,,
\end{equation}
where $\delta_{s}$ is the exact distortion of the background Kerr
solution and measuring the relative difference of the shadow at $\psi =0$
and $\psi = \pi$. On the other hand, $\delta_{s,*}$ is given by either
$\delta_{s,_{\rm I}}-\delta_{s,_{\rm III}}$ or
$\delta_{s,_{\rm HM}}$. 

Figure~\ref{distribution} shows the distributions of the errors computed
in this way for the four different possible definitions of the distortion
parameters, with the black line referring to the distortion parameter in
Eq. (\ref{delthioki}), and the red, blue and light-blue lines referring
to the definitions (\ref{deltadef1}), (\ref{deltadef2}), and
(\ref{deltadef3}), respectively. Note that the values of the probability
densities distributions are reported in such a way that
\begin{equation}
\frac{1}{N_{\rm tot}}\int^{\infty}_{-\infty} dn =
\frac{1}{N_{\rm tot}}\int^{\infty}_{-\infty}
\left(\frac{dn}{d\epsilon_*} \right)
d \epsilon_* = 1 \,.
\end{equation}

The distributions reported in Fig. \ref{distribution} are rather
self-explanatory. Clearly, all the different definitions are centred on
$\epsilon_*=0$, indicating that on average they provide a good
measurement of the distortion. On the other hand, the variance of the
different distribution is rather different. Overall, the distortion
parameters $\delta_{s,_{\rm I}}-\delta_{s,_{\rm III}}$ have comparable
variances, with a slightly smaller variations for the definition
$\delta_{s,_{\rm I}}$. However, the variance of the distortion parameter
for $\delta_{s,_{\rm HM}}$ is almost twice as large as the others and it
essentially spans the $5\%$ variation that we introduce in the random
distortion $\Delta_c$. These results are rather reassuring as they
indicate that new definitions are not only accurate, but also robust with
respect to random white noise. Furthermore, they appear to be superior to
other distortion measurements suggested in the past.

As a final remark we note that the introduction of the perturbations in
the expansion (\ref{perturbed}) also has the effect of changing the
position of the effective centre of the shadow and hence the values of
$\vec{\boldsymbol{R}}_0$ and $\psi_0$ [\cf Eqs. \eqref{eq:center_cont}
  and (\ref{psizero})]. Fortunately, such variations represent only a
high-order error, which is much smaller than those measured by the
distortion parameters, with maximum measured variance of the order of
$10^{-4}$. As a result, the distortions reported in
Fig. ~\ref{distribution} are genuine measurements of the shadow and not
artefacts introduced by the changes in the effective centres.

\section{Conclusion}
\label{conclusion}

The radio-astronomical observations of the shadow of a black hole would
provide convincing evidence about the existence of black holes. Further,
the study of the shadow could be used to learn about extreme gravity near
the event horizon and determine the correct theory of gravity in a regime
that has not been explored directly so far. A number of different
mathematical descriptions of the shadow have already been proposed, but
all make use of a number of simplifying assumptions that are unlikely to
be offered by the real observational data, \eg the ability of knowing
with precision the location of the centre of the shadow.

To remove these assumptions we have developed a new general and
coordinate-independent formalism in which the shadow is described as an
arbitrary polar curve expressed in terms of a Legendre expansion. Our
formalism does not presume any knowledge of the properties of the shadow
and offers a number of routes to characterize the properties of the
curve. Because our definition of the shadow is straightforward and
unambiguous, it can be implemented by different teams analysing the same
noisy data.

The Legendre expansion used in our approach converges exponentially
rapidly and we have verified that a number of coefficients less than ten
is sufficient to reproduce the shadow with a precision of one part in
$10^5$, both in the case of a Kerr black hole with spin parameter of
$a/M=0.99$, and in the case of Bardeen and Kerr-Taub-NUT black holes with
large magnetic charges and NUT parameters. Furthermore, the use of a
simple Legendre expansion has allowed us to introduce three different
definitions of the distortion of the shadow relative to some reference
circles. The comparison of the different definitions has allowed us to
determine which of them is best suited to capture the complex structure
of the shadow and its distortions.

Finally, again exploiting the advantages of the Legendre expansion, we
were able to simulate rather simply the presence of observational random
errors in the measurements of the shadow. Constructing a large number of
synthetically perturbed shadows, we have compared the abilities of the
different parameters to measure the distortion in the more realistic case
of a noisy shadow. Overall, we have found that our new definitions have
error distributions with comparable variances, but also that these are
about a factor of 2 smaller than the corresponding variance measured
when using more traditional definitions of the distortion. Given these
results, the approach proposed here could be used in all those studies
characterizing the shadows of black holes as well as in the analysis of
the data from experimental efforts such as EHT and BHC.

\section*{Acknowledgements}

It is a pleasure to thank A. Grenzebach, Y. Mizuno, Z. Younsi, and
A. Zhidenko for useful discussions and comments. We are also grateful to the referee, D. Psaltis, for comments and suggestions that have improved the presentation. This research was
partially supported by the Volkswagen Stiftung (Grant 86 866) and by the
ERC Synergy Grant ``BlackHoleCam -- Imaging the Event Horizon of Black
Holes'' (Grant 610058). AAA and BJA are also supported in part by the
project F2-FA-F113 of the UzAS and by the ICTP through the projects
OEA-NET-76, OEA-PRJ-29. AAA and BJA thank the Institut f{\"u}r
Theoretische Physik, Goethe Universit\"{a}t  for warm hospitality during their stay in Frankfurt.

\bibliographystyle{mn2e}
\bibliography{manuscript}

\appendix

\section{Graphical representation of distortion parameters: Kerr black holes}

\begin{figure*}
\begin{center}
  \includegraphics[width=0.31\textwidth]{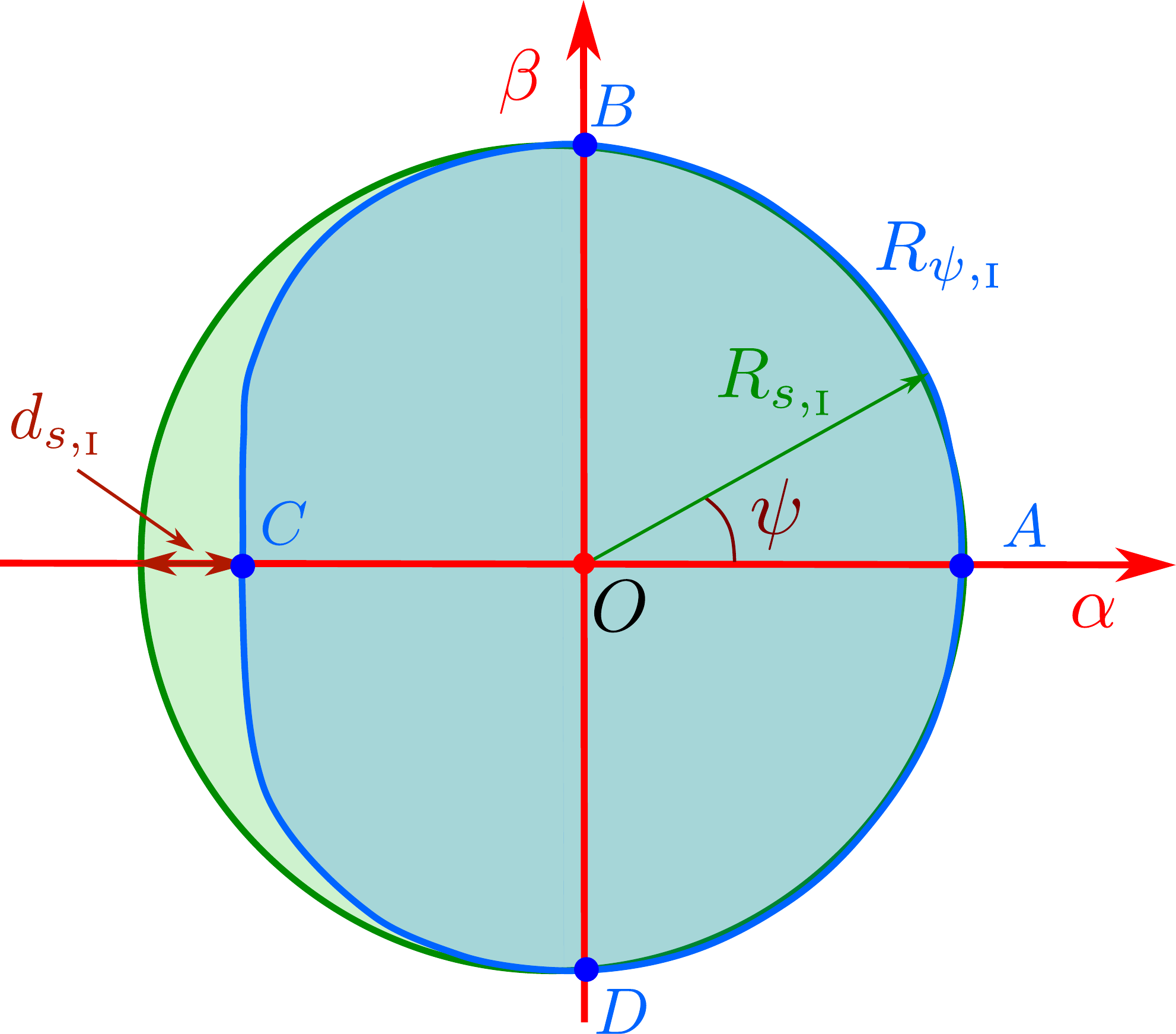}
  \hskip 0.5cm
  \includegraphics[width=0.31\textwidth]{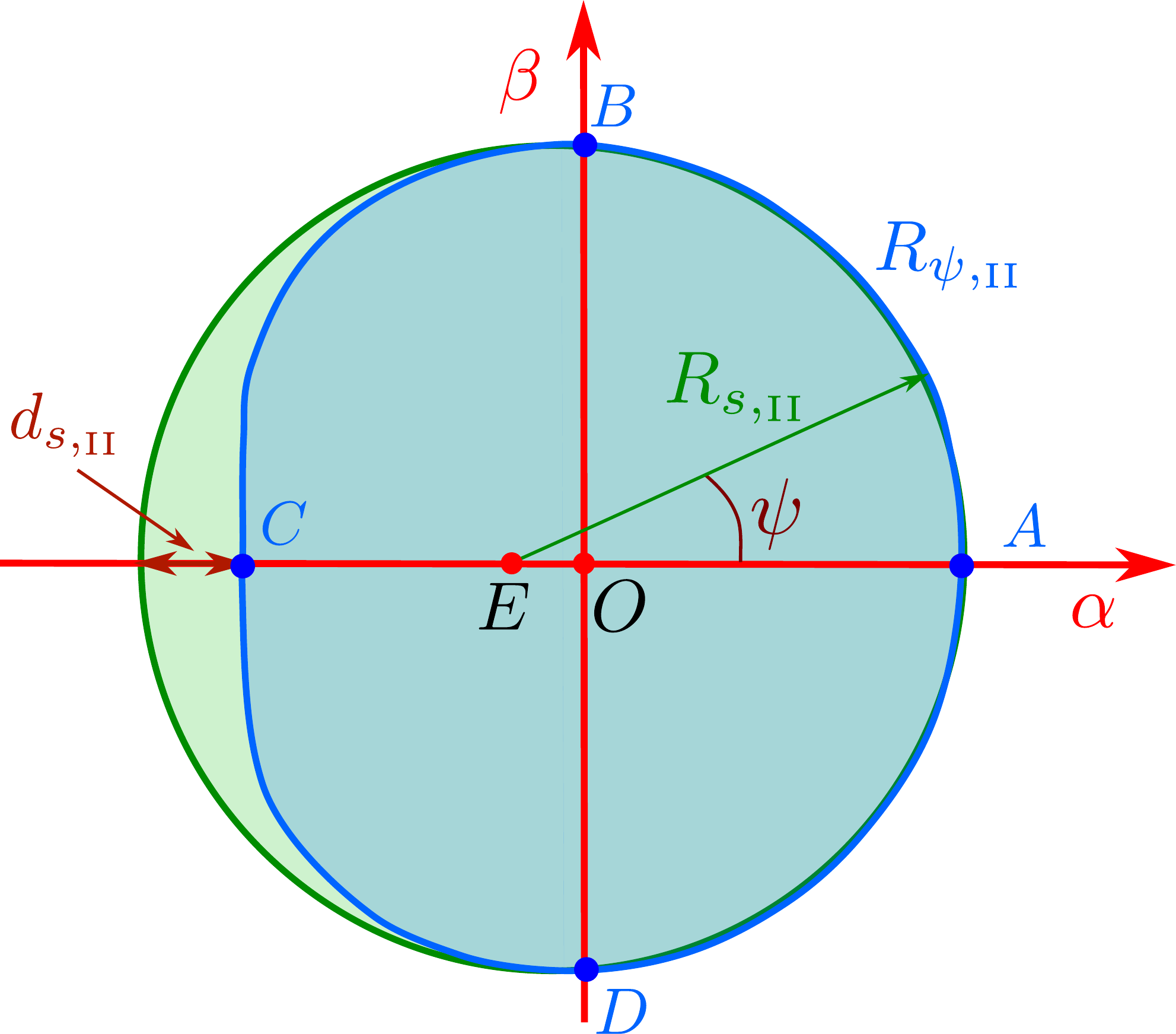}
  \hskip 0.5cm
  \includegraphics[width=0.31\textwidth]{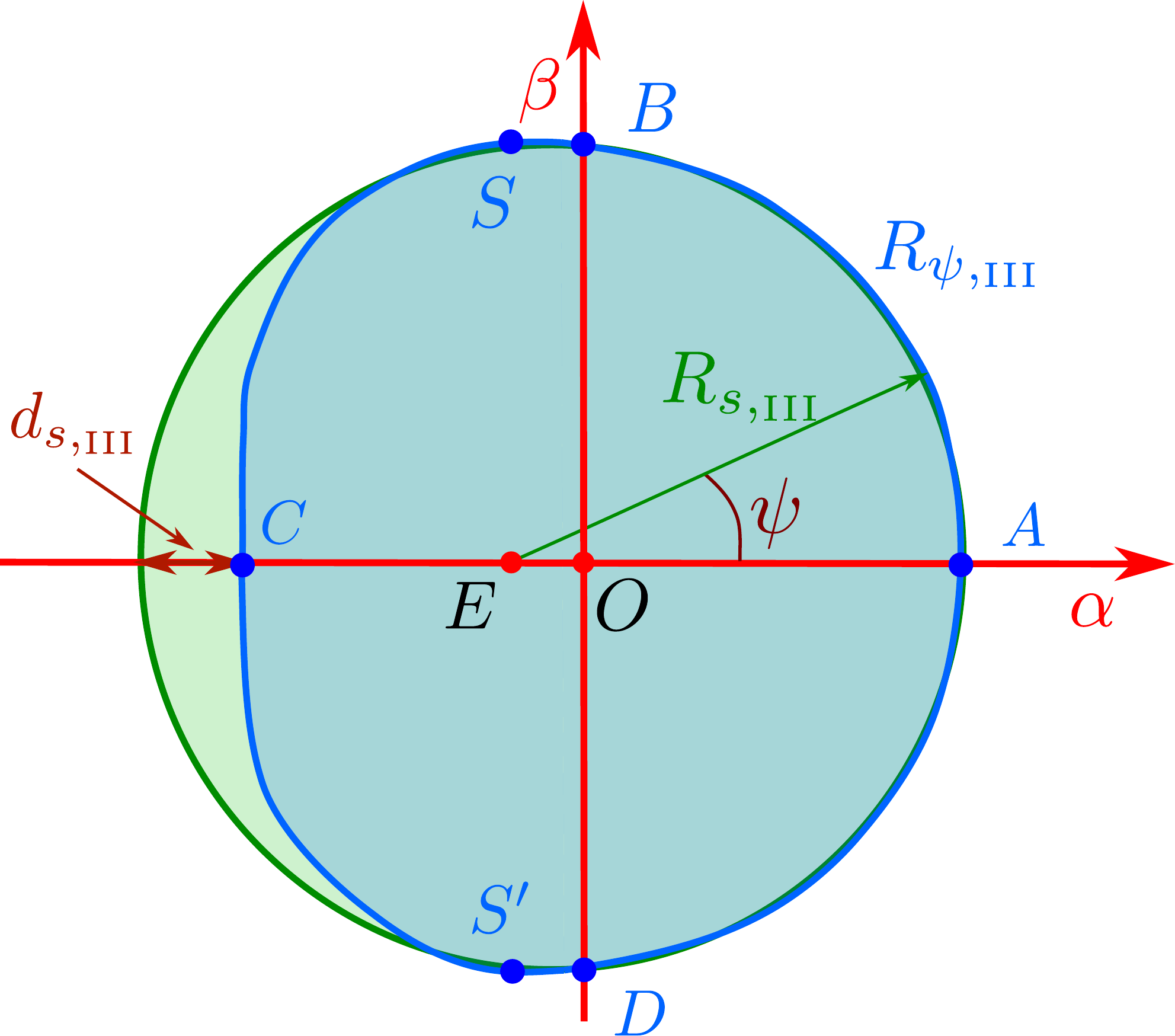}
\end{center}
\caption{Schematic representations of the distortion parameters I, II and
  III when applied to the shadow of a Kerr black hole. The left-hand panel
  refers to the definition I, where the centre of the reference circle as
  well as the points $B$ and $D$ (which are not slope points) are on the
  coordinate axis $\beta$. The middle panel refers instead to definition
  II and in this case the centre of the reference circle $E$ is displaced
  along the $\alpha$ axis. Finally, the right-hand panel shows the
  representation of the definition III, where we consider the reference
  circle passing through the point $A$ and the slope points $B$ and $D$
  are not on the $\beta$ axis. The centre of the reference circle $E$ is
  also displaced and not exactly at the centre of the coordinate system.}
\label{fig:appendix}
\end{figure*}

To help in the visualization of the different distortion parameters
introduced in Section \ref{dist_params}, we here present their graphical
representation when they are applied to the shadow of a Kerr black hole
shadow with spin parameter $a^*=0.98$. The left-hand panel of
Fig.~\ref{fig:appendix} represents the distortion parameter I in
Sect.~\ref{dist_params}, where the reference circle is defined in a such
a way that the points $B, D$ and the centre of the reference circle are
all on the vertical axis $\beta$.  Hence, the difference of the radial
coordinates of the point $C$ of the shadow boundary and of the left-hand point
of the circle intersecting the axis $\alpha$ corresponds to the
distortion parameter $d_{s,{\rm I}}$.

Similarly, the middle panel of Fig.~\ref{fig:appendix} refers to the
distortion parameter II, where the reference circle passes through the
points $A$ and $B$, which are on the axis $\beta$. The centre of the
reference circle is on the point $E$ and does not coincide with the
centre of the coordinate system. The position of the reference circle
centre $E$ is instead defined by Eq.~(\ref{retwo}). The difference of the
radial coordinates of the point $C$ of the shadow and of the left-hand point
of the circle intersecting with the axis $\alpha$ corresponds to the
distortion parameter $d_{s,{\rm II}}$.

Finally, the right-hand panel of Fig.~\ref{fig:appendix} corresponds to the
distortion parameter III, where the reference circle passes through the
slope points $S, S'$, and the right-hand point $A$ of the shadow intersecting
the axis $\alpha$. The positions of the slope points are defined by
solving Eq.~(\ref{slope_eq}). The centre of the reference circle does not
coincide with the origin of the coordinate system and its position is
defined by Eq.~(\ref{rrthree}). The difference of the radial coordinates
of the point $C$ of the shadow and of the left-hand point of the circle
intersecting the axis $\alpha$ corresponds to the distortion parameter
$d_{s,{\rm III}}$.

Note that the radius of the reference circle radius depends on the
definition used for the distortion parameter; hence the distortion
parameters are also different for each definition.

\label{lastpage}

\end{document}